\documentclass[11pt]{article}

\usepackage{amsfonts}
\usepackage{amssymb}
\usepackage{amsmath}
\usepackage{amsthm}
\usepackage{graphicx}

\def\be{\begin{equation}}
\def\ee{\end{equation}}
\def\bc{\begin{center}}
\def\ec{\end{center}}

\begin{document}

\hspace{13.9cm}1

\ \vspace{20mm}\\

{\LARGE \noindent Decorrelation by recurrent inhibition in heterogeneous neural circuits}

\ \\
{\bf \large Alberto Bernacchia$^{\displaystyle 1}$}\\
{\bf \large Xiao-Jing Wang$^{\displaystyle 1}$}\\
{$^{\displaystyle 1}$Department of Neurobiology, Yale University, 333 Cedar Street, New Haven CT.}\\

%


\thispagestyle{empty}
\markboth{}{NC instructions}
\ \vspace{-0mm}\\
%
\begin{center} {\bf Abstract} \end{center}
The activity of neurons is correlated, and this correlation affects how the brain processes information.
We study the neural circuit mechanisms of correlations by analyzing a network model characterized by strong and heterogeneous interactions: excitatory input drives the fluctuations of neural activity, which are counterbalanced by inhibitory feedback.
In particular, excitatory input tends to correlate neurons, while inhibitory feedback reduces correlations.
We demonstrate that heterogeneity of synaptic connections is necessary for this inhibition of correlations.
We calculate statistical averages over the disordered synaptic interactions, and we apply our findings to both a simple linear model and to a more realistic spiking network model.
We find that correlations at zero time-lag are positive and of magnitude $K^{-\frac{1}{2}}$, where $K$ is the number of connections to a neuron.
Correlations at longer timescales are of smaller magnitude, of order $K^{-1}$, implying that inhibition of correlations occurs quickly, on a timescale of $K^{-\frac{1}{2}}$. 
The small magnitude of correlations agrees qualitatively with physiological measurements in the Cerebral Cortex and Basal Ganglia.
The model could be used to study correlations in brain regions dominated by recurrent inhibition, such as the Striatum and Globus Pallidus.

\section{Introduction}
Simultaneous measurements of the activity of multiple neurons have shown significant correlations, and this observation has stimulated the debate on whether and how correlations contribute to neural computation. 
In principle, correlations allow robust signal processing, because redundancies across neurons can be exploited to separate the signal from the noise \cite{abbott99, panzeri99}. 
Experimental studies of the Cerebral Cortex suggest that correlations improve decoding of stimuli \cite{graf11}, but it remains unclear whether a parsimonious decoder should rely on correlations \cite{averbeck03}.
A challenge to this hypothesis is the observation that correlations are reduced when animal subjects are actively engaged in discrimination \cite{cohen08, cohen09}, and even when they simply start a movement \cite{poulet08}.
In addition, neurons with similar responses to stimuli show higher correlations \cite{zohary94, lee98, maynard99, bair01, constantinidis02, averbeck03, romo03, kohn05, smith08, huang09, ecker10, komiyama10}, implying that coding of stimuli should be worsened by correlations \cite{abbott99, panzeri99, sompolinsky01, wilke02, averbeck06, gutniski08}.
Another caveat is that the neural code is largely unknown, and if the "noise" measured in physiological studies encodes some signal then any correlation would decrease the available information \cite{nadal94}. 

Besides the possible function of correlations in signal and information processing, their physiological causes remain unclear.
It has been shown that the correlation between nearby neurons is driven by their correlated synaptic input \cite{lampl99, poulet08}.
However, a quantitative understanding of the circuit mechanisms regulating correlations between cortical cells is still missing, and the goal of this study is to determine the dependence of correlations on different properties of the neural circuitry. 
The measured correlation between neurons depends on different factors and varies across studies \cite{cohen11}: it increases with the proximity of neuron pairs \cite{maynard99, constantinidis02, smith08, ecker10, komiyama10}, their activity \cite{delarocha07} and the temporal window on which action potentials are counted \cite{bair01, reich01, constantinidis02, averbeck03, kohn05, smith08, huang09, mitchell10}.
Fig.1 shows the correlation measured in eight different studies as a function of temporal window for spike counts.
Results vary, although correlations are generally found positive and of small magnitude, both in the Cortex and the Basal Ganglia \cite{raz00}.

Previous modeling studies of neural circuits have found that the mean correlation between neurons is small, of the order of $N^{-1}$, where $N$ is the number of neurons in the network.
Small correlations have been observed, not surprisingly, in networks characterized by weak connection strengths \cite{ginzburg94, bernacchia07}.
More surprisingly, the same result has been obtained in the case of strong connections, such as the high-conductance state \cite{destexhe03, vanvreeswijk96}, provided that the network includes a strong inhibitory feedback \cite{renart10, hertz10, tetzlaff12}.
Here we provide an analytical study of correlations in a simple linear model, and we apply our findings to predict correlations in a more realistic spiking network model.
We confirm both the observed small correlation, and the crucial effect of the inhibitory feedback in reducing it.
In addition, we study the effect of the heterogeneity of connection strengths by using random matrix theory and a diagrammatic formalism, and we show that inhibition of correlations crucially depends on such heterogeneity.
The model can be compared to brain regions dominated by recurrent inhibition, such as the Striatum and Globus Pallidus.

We find that correlations at zero time-lag are of magnitude $K^{-\frac{1}{2}}$, where $K$ is number of connections received by a neuron, while correlations of the activity integrated across time is of order $K^{-1}$, suggesting that inhibition of correlations operates on a timescale of $K^{-\frac{1}{2}}$.
These results are consistent with previous modeling studies, suggesting that a linear approximation is adequate to predict correlations in more realistic spiking models \cite{renart10}.
In addition, our findings highlight the difference between the effect of the number of neurons $N$ versus the number of connections $K$ on correlations.
The small correlations predicted by this and previous modeling studies qualitatively match the small correlations observed in neurons of the Cerebral Cortex and Basal Ganglia.

\begin{figure}[h!]
\begin{center}
\includegraphics[width=4in]{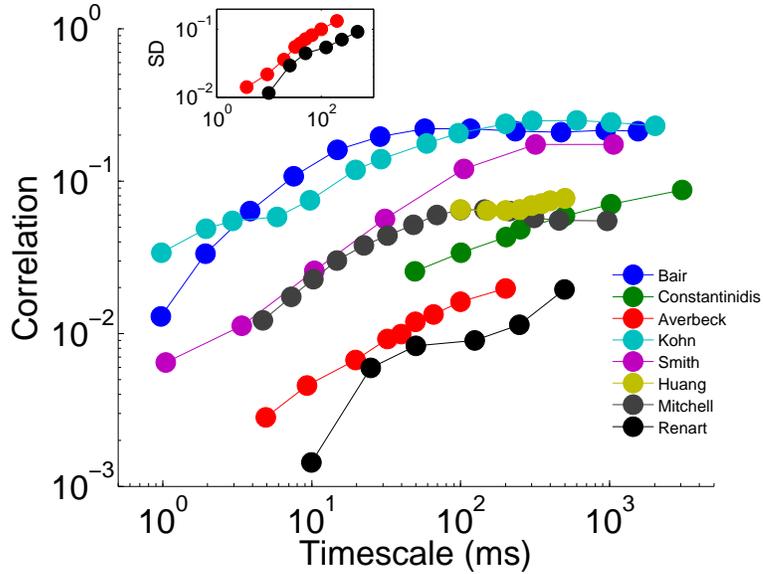}
\end{center}
\caption{Mean correlation across neuron pairs plotted vs the length of the time window used to count action potentials.
Re-plotted from eight experimental studies of the cortex (legend).
Two studies provided not only the mean correlation but also the Standard Deviation (SD, inset).
}
\label{eightstudies}
\end{figure}

\section*{Methods}
We consider both a linear model and a more realistic spiking network model.
In both models, we consider a neural circuit of $N$ neurons, receiving input from $N_{ext}$ external neurons, where each neuron integrates the signal from other neurons weighted by the synaptic connection strength.

The dynamics of the linear model is described by the equation

\begin{equation}
\label{xdyn1}
\tau\frac{dx_i(t)}{dt}=-x_i(t)+\sum_{j=1}^N{G_{ij}x_j(t)}+\sum_{j=1}^{N_{ext}}G^{ext}_{ij}x_j^{ext}(t)
\end{equation}
where $x_i$ is the activity of neuron $i$ in the local circuit, and $G_{ij}$ is the strength of the synaptic connection from neuron $j$ to neuron $i$. 
The external (feed-forward) input to the circuit is provided by the activities $x_j^{ext}$, and the synaptic connection from the $j$-th external neuron to the $i$-th local neuron is given by the strength $G^{ext}_{ij}$. 
All neuronal activities evolve in time, while the connectivity matrices $G$ and $G_{ext}$ are fixed.

\begin{figure}[h!]
\begin{center}
\includegraphics[width=4in]{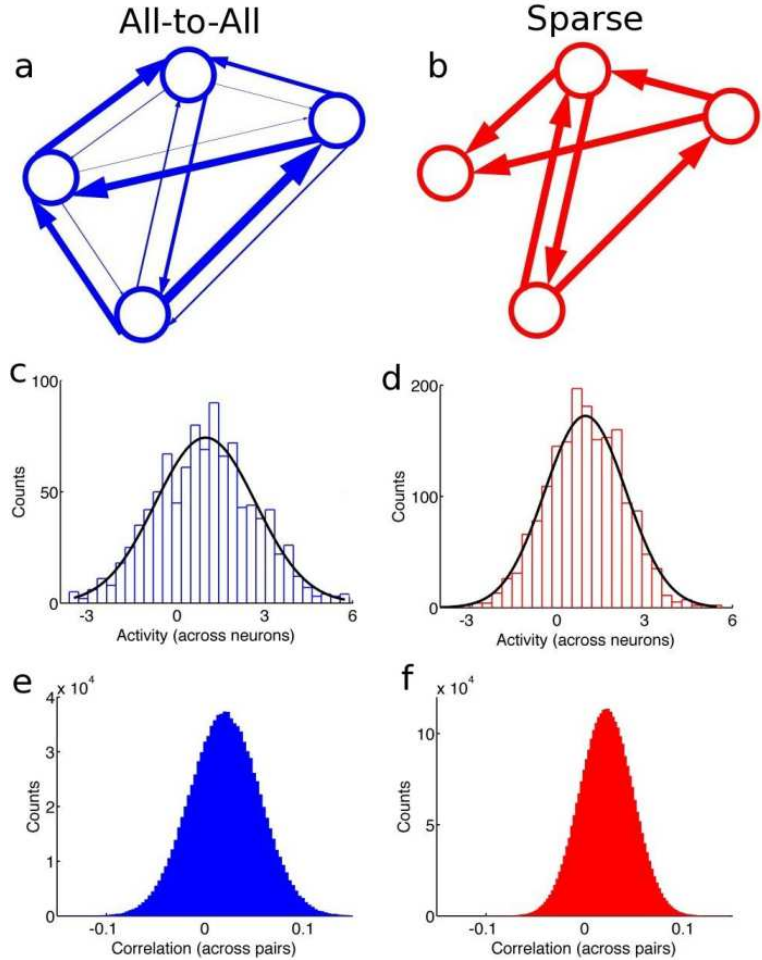}
\end{center}
\caption{
Scheme of the network and distribution of activity and correlations.
We study two network architectures: All-to-All connectivity with random strengths (a), and Sparse random connections of fixed strength (b).
Connection strength is illustrated by the thickness of edges.
The distribution of activity across neurons (c,d) and the distribution of correlations across neuron pairs (e,f) are Gaussian for both types of networks ((c,e) for the All-to-All network, (d,f) for the Sparse network).
The parameters used are: $g=g_{ext}=1$, $\overline{x}_{ext}=\overline{\Delta x_{ext}^2}=1$, $K_{ext}=K$.
For the All-to-All network: $K=1000$, $\lambda_{ext}=1$, $\lambda=1/\sqrt{2}$.
For the sparse network: $K=880$ and $k_{ext}=k=1/2$, which correspond to $\lambda=\lambda_{ext}=1/\sqrt{2}$.}
\label{Fig_net}
\end{figure}

We define the average number of local connections received by a neuron as $K$, and the external connections as $K_{ext}$.
We assume that the connectivity matrices are random, which makes the network akin to a "disordered" system, which is characterized by a random but fixed substrate.
We consider two scenarios (represented schematically in Fig.\ref{Fig_net}a,b): 
\newline
\newline
\parbox{\textwidth}{1) The network is fully connected ($K=N$, $K_{ext}=N_{ext}$) with random connection strengths (All-to-All, Fig.\ref{Fig_net}a), characterized by a  Gaussian distribution. 
The mean and variance of matrix elements are determined by the parameters $g$ and $\lambda$ for the local connections, $g_{ext}$ and $\lambda_{ext}$ for the external connections:}
\begin{equation}
\label{synmat1fc}
\left<G_{ij}\right>=-g/\sqrt{N}\;\;\;\;\;\;\;\left<\Delta G_{ij}^2\right>=\lambda^2/N
\end{equation}\begin{equation}
\label{synmat2fc}
\left<G^{ext}_{ij}\right>=g_{ext}/\sqrt{N_{ext}}\;\;\;\;\;\;\;\left<\Delta {G^{ext}_{ij}}^2\right>=\lambda_{ext}^2/N_{ext}.
\end{equation}
\newline
\parbox{\textwidth}{2) The network is sparse, only a fraction of connections exists ($k=K/N$, $k_{ext}=K_{ext}/N_{ext}$), the others are set to zero (Sparse, Fig.\ref{Fig_net}b). Connections are selected at random but of constant strength, equal to $-g/\sqrt{K}$ for recurrent connections and $g_{ext}/\sqrt{K_{ext}}$ for external connections. The distribution is Bernouillian; The mean and variance of matrix elements are:}
\begin{equation}
\left<G_{ij}\right>=-kg/\sqrt{K}\;\;\;\;\;\;\;\left<\Delta G_{ij}^2\right>=k(1-k)g^2/K
\end{equation}\begin{equation}
\left<G^{ext}_{ij}\right>=k_{ext}g_{ext}/\sqrt{K_{ext}}\;\;\;\;\;\;\;\left<\Delta {G^{ext}_{ij}}^2\right>=k_{ext}(1-k_{ext})g_{ext}^2/K_{ext}.
\end{equation}
\newline
Angular brackets denote average over the matrix distribution, $\Delta$ indicates variation around the mean.
We adopt a single notation for either case, All-to-All or Sparse network, by defining the mean and variance of matrix elements and their scaling with $K$, $N$: 

\begin{equation}
\label{synmat1}
\left<G_{ij}\right>=-kg/\sqrt{K}\;\;\;\;\;\;\;\left<\Delta G_{ij}^2\right>=\lambda^2/N
\end{equation}
\begin{equation}
\label{synmat2}
\left<G^{ext}_{ij}\right>=k_{ext}g_{ext}/\sqrt{K_{ext}}\;\;\;\;\;\;\;\left<\Delta {G^{ext}_{ij}}^2\right>=\lambda_{ext}^2/N_{ext}.
\end{equation}
In the All-to-All network, $k=1$ and $K=N$.
In the Sparse network, for convenience of notation we use the parameters $\lambda^2=g^2(1-k)$ and $\lambda_{ext}^2=g_{ext}^2(1-k_{ext})$.
The mean connection is negative for $G$ (inhibitory) and positive for $G_{ext}$ (excitatory), since $g$ and $g_{ext}$ are positive.
Note that the connections are "strong" in the sense that the magnitude of the excitatory and inhibitory input to each neuron, which is of order $\sqrt{K}$, is much larger than their sum (which is of order one, see Results).

Theoretical analysis considers also the case in which local connections can be either excitatory or inhibitory, with two separate populations of excitatory and inhibitory neurons.
In general, the analysis considers the case in which the mean and variance of the synaptic strength depends on the pre-synaptic neuron.
We discuss in Appendix 2 that all theoretical results hold provided that the parameters $g$ and $\lambda^2$ are substituted by the means across pre-synaptic neurons.
However, we do not show results of simulations for that case since that is outside the scope of the present work.  

We assume that the external activity $\textbf{x}_{ext}(t)$ is a stochastic process uncorrelated in both space and time, i.e. a white noise characterized by mean $\overline{x_i^{ext}(t)}=\overline{x}_{ext}$ and covariance $\overline{\Delta x_i^{ext}(t)\Delta x_j^{ext}(t')}=\overline{\Delta x_{ext}^2}\delta_{ij}\delta(t-t')$ 
(overline denotes the average over different realizations of the noise, and $\delta$ denotes either the discrete Kronecker or continuous Dirac function).
Therefore, Eq.(\ref{xdyn1}) corresponds to a Ornstein-Uhlenbeck stochastic process \cite{gardiner}.

We test theoretical results by running numerical computer simulations of the linear model.
We simulate the dynamics of Eq.(\ref{xdyn1}) with a simple Euler integration method, where each simulation runs for 200,000 time steps and each time step is 0.002 $\tau$. 
For each set of parameter values, we use a single realization of the external input noise and a single realization of the random connectivity matrix. 
Since a simulation runs for a long time and the network is composed by a large number of neurons, we do not expect those specific realizations to affect the results significantly (in other words, we expect the system to be ergodic and self-averaging). 
We calculate sample mean and covariance by averaging across all time steps of a simulation. 
The correlation is calculated for each neuron pair by the standard Pearson's formula. 
Finally, we calculate the spatial mean and variance of those quantities across neurons (for the temporal mean) or across neuron pairs (for the covariance and correlation). 
We also use a semi-analytic control by applying the spatial mean and variance on, instead of the full simulation, the theoretical results following the average over temporal noise but preceding the average over spatial noise, namely Eqs.(\ref{thermalx}),(\ref{C_OU}).
The corresponding results are represented by filled symbols in the figures, while results of the full simulations are represented by open symbols.

The spiking network is defined by a current-based Integrate and Fire model. Its dynamics is described by the equations

\begin{equation}
\label{Idyn}
\tau\frac{dI_i(t)}{dt}=-I_i(t)+\sum_{j=1}^N{G_{ij}S_j(t)}+\sum_{j=1}^{N_{ext}}G^{ext}_{ij}S_j^{ext}(t)
\end{equation}

\begin{equation}
\label{Vdyn}
C_m\frac{dV_i(t)}{dt}=-g_m(V_i(t)-V_L)+I_i(t)
\end{equation}
These equations are integrated using a simple Euler method with a time step $dt=0.02ms$. 
Eq.(\ref{Idyn}) is similar to Eq.(\ref{xdyn1}) of the linear model, and describes the dynamics of the total current $I_i$ received by neuron $i$, both the external excitatory and the recurrent inhibitory input (both types of input are integrated according to the same time constant $\tau$).
The matrices describing the synaptic strengths, $G$ and $G_{ext}$, are defined in the same way as in the case of the linear model, in the fully connected case (see Eqs.(\ref{synmat1fc},\ref{synmat2fc})), although in the spiking model those matrices are given in units of $8\;nA\cdot ms$.
In those units, parameters are: $g=1$, $\lambda=0.5$, $g_{ext}=1$, $\lambda_{ext}=0.58$.
The variable $S_i(t)$ describes whether neuron $i$ emits an action potential at time $t$ or not, respectively, $S_i(t)=1/dt$ or $S_i(t)=0$.
Eq.(\ref{Vdyn}) describes the dynamics of the membrane potential $V_i$ of neuron $i$, which integrates linearly the total current according to the capacitance $C_m$ and conductance $g_m$ of the membrane, where $V_L$ is the resting potential.
If the membrane potential $V_i$ exceeds the threshold potential $V_{th}$ at time $t$, it is set to the reset potential $V_{rs}$ and an action potential is emitted ($S_i(t)=1/dt$).
The variable $S_j^{ext}(t)$ describes the action potentials emitted by the external neurons. 
Their activity is modeled by a Poisson process characterized by an emission rate $\phi_{ext}$, which is constant in time and equal for all external neurons.
Parameters used in simulations are: $\tau=10ms$, $V_{th}=-50mV$, $V_{rs}=-70mV$, $V_L=-70mV$, $\phi_{ext}=50Hz$, $C_m=0.4nF$, $g_m=20nS$ (the time constant of the membrane potential is $C_m/g_m=20ms$).

We run $20$ simulations for different values of the network size, from $N=50$ to $N=1000$, with all other parameters fixed, each simulating $20s$ of network activity ($10^6$ time steps).
For a given network size, $N=200$, we run additional $5$ simulations at different values of the external input, from $\phi_{ext}=45Hz$ to $\phi_{ext}=55Hz$.
We use those $5$ simulations to determine the change of the total current as a function to the change in $\phi_{ext}$. This change is approximately linear, and is quantified in terms of the four statistics studied in this work, namely the mean, the spatial variance, the temporal variance and the covariance (see e.g. Eqs.(\ref{mx}),(\ref{s2x}),(\ref{ondiag2}),(\ref{offdiag})).
All statistics are calculated with respect to the currents measured at the reference external input,  $\phi_{ext}=50Hz$.
For example, the spatial variance is calculated by recording the steady current for each neuron at the reference value, and looking at the distribution across neurons of the difference between the reference current and the steady currents measured for the other values of the external input.
Linear regression is applied to fit the linear change of the four statistics as a function of the external input, and the "effective" parameters ($g, \lambda, g_{ext}, \lambda_{ext}$) are determined by inverting the equations of the four statistics given by the linear model, Eqs.(\ref{mx}),(\ref{s2x}),(\ref{ondiag2}),(\ref{offdiag}), where $\overline{x}_{ext}$ and $\overline{\Delta x_{ext}^2}$ are the mean and variance of the change in external rate ($\overline{x}_{ext}=(\phi_{ext}-50Hz)$ and $\overline{\Delta x_{ext}^2}=(\phi_{ext}-50Hz)/\tau$).
The effective parameters are used in Eq.(\ref{corr}) to predict correlations in the spiking model at variable network size.

\section*{Results}

We study neural activity and correlations among neurons in a heterogeneous neural circuit model.
Local recurrent connections are dominated by inhibition, while external feed-forward projections are excitatory.
Results are shown for a simple linear model and, at the bottom of the section, we also include simulations of a more realistic spiking network model.
For the linear model we show the results of both theory and simulations, and we conclude by showing that the theory developed for the simple linear model can be used to predict correlations in the spiking network.

We consider two types of circuits, All-to-All connectivity with random strengths (Fig.\ref{Fig_net}a) and Sparse random connections of fixed strengths (Fig.\ref{Fig_net}b).
Results are displayed in a single notation for either case (see Methods).
Fig.\ref{Fig_net}c,d shows the distribution of activity across neurons, and Fig.\ref{Fig_net}e,f shows the distribution of correlations across neuron pairs.
The purpose of this work is to describe how the mean and variance of those distributions depend on the parameters of the neural circuit.
The activity values $x$ are interpreted as deviations from a steady state of the input currents to each neuron, around which the neural dynamics is approximately linear.
If we denote the steady current as $I_0$, the input current is equal to $I=I_0+x$.
As long as the linear approximation is valid, the correlations observed in the model are insensitive to the nature of the steady state (i.e. to the value of $I_0$).

\begin{figure}[h!]
\begin{center}
\includegraphics[width=4in]{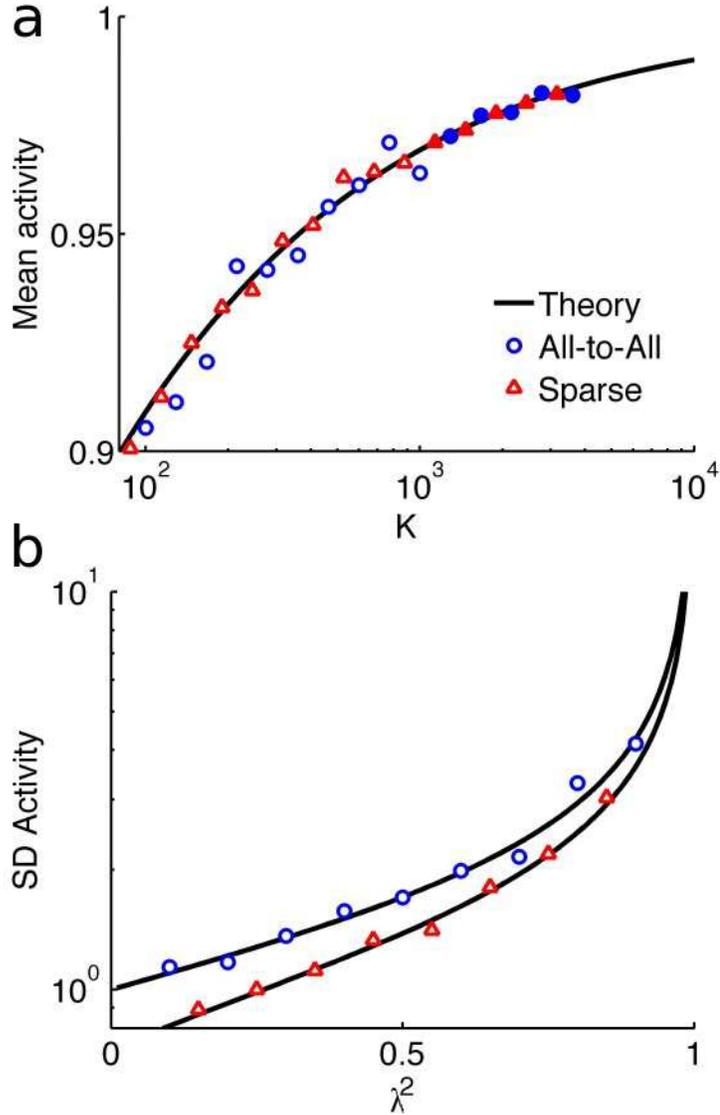}
\end{center}
\caption{
The mean activity has a mild dependence on the number of connections $K$ (a); Its Standard Deviation (SD) strongly depends on the heterogeneity of connections $\lambda$ (b).
Analytical results (lines) are obtained from Eq.(\ref{mx}) in panel a and Eq.(\ref{s2x}) in panel b.
Simulation results are shown for the All-to-All (blue circles) and Sparse (red triangles) network.
Open symbols show simulations of the neural dynamics, Eq.(\ref{xdyn1});
Filled symbols show numerical evaluation of Eq.(\ref{thermalx}).
The parameters used are, in both panels, $g=g_{ext}=1$, $\overline{x}_{ext}=\overline{\Delta x_{ext}^2}=1$.
In panel b, $K_{ext}=K=500$.
For the All-to-All network: $\lambda_{ext}=1$, and $\lambda=1/\sqrt{2}$ in panel a.
For the Sparse network: $k_{ext}=k=1/2$ in panel a (which correspond to $\lambda=\lambda_{ext}=1/\sqrt{2}$), while $k_{ext}=k$ is varied in panel b according to the value of $\lambda$ (see Methods).}
\label{Fig_act}
\end{figure}

Due to the linearity of the model, all quantities of interests can be simply calculated;
The novel contribution of this work is averaging those quantities over the randomness of the connectivity matrix.
Because connections are heterogeneous, different neurons have a different activity, and we compute the sample mean across neurons in order to obtain the spatial average.
If the number of neurons $N$ is large, this is independent on the specific realization of the connectivity, therefore we perform its average over the distribution of connections, and we obtain (see Eq.(\ref{mxA}) in Appendix 1; Angular brackets denote averaging over the random connectivity, overline denotes temporal average)

\begin{equation}
\label{mx}
\left<\overline{x}\right>=\left<\frac{1}{N}\sum_{i=1}^N\overline{x_i}\right>=\frac{g_{ext}\sqrt{K_{ext}}}{1+g \sqrt{K}}\,\overline{x}_{ext}
\end{equation}
The numerator of this expression is equal to the mean excitatory input received by a neuron, $g_{ext}\sqrt{K_{ext}}\;\overline{x}_{ext}$, while the denominator expresses the recurrent inhibition, whose total post-synaptic strength is $g\sqrt{K}$.
Therefore, the strong recurrent inhibition counterbalances the large excitatory input and determines a relatively low activity, regardless of the network size.
Note that the number of local and external connections, $K$ and $K_{ext}$, are both large, but they tend to balance in the expression above.
Fig.\ref{Fig_act}a shows an example of mean activity as a function of the number of connections;
The mean activity is rather insensitive to the number of connections, which are taken equal to the external ones in each simulation ($K=K_{ext}$).
The analytical result, Eq.(\ref{mx}), agrees with numerical simulations of the linear dynamics, in both the All-to-All and the Sparse networks.

Different neurons have different connections and therefore different activity, and the extent to which the activity varies from neuron to neuron is determined by the spatial variance.
We calculate this quantity by taking the sample variance across neurons and averaging over the random connectivity, and we obtain (see Eq.(\ref{s2xA}) in Appendix 1)

\begin{equation}
\label{s2x}
\left<\Delta\overline{x}^2\right>=\left<\frac{1}{N}\sum_{i=1}^N\Delta\overline{x_i}^2\right>=\frac{1}{1-\lambda^2}\left[\left<\overline{x}\right>^2\lambda^2+\overline{x}_{ext}^2\lambda_{ext}^2\right]
\end{equation}
The spatial variance of neural activity increases with the network heterogeneity, expressed by $\lambda$ and $\lambda_{ext}$ for, respectively, the recurrent and external connections.
Increasing the heterogeneity of connections increases the differences in the total input between neurons and therefore in their activities.
The spatial variance is also proportional to the mean activity, local $\left<\overline{x}\right>$ and external $\overline{x}_{ext}$.
Furthermore, increasing the heterogeneity of recurrent connections leads to a divergence of the spatial variance, when $\lambda^2$ approaches one.
In this case the state $x=0$ destabilizes, and the linear approximation fails (see Appendix 1).
Fig.\ref{Fig_act}b shows an example of the spatial variance as a function of the variability of the recurrent connections.
The analytical result, Eq.(\ref{s2x}), agrees with numerical simulations of the linear dynamics, in both the All-to-All and the Sparse network.

After looking at the mean and spatial variability of neural activity, we turn to the main theme of our work, the analysis of temporal variability and correlations.
The activity of each neuron fluctuates in time, due to the fluctuating input, and those temporal fluctuations may be correlated since different neurons receive shared input.
We study temporal variability and correlated fluctuations by calculating the covariance matrix, in particular the instantaneous covariance, at zero time lag.
This is defined as

\begin{equation}
Q_{ij}=\overline{\Delta x_i\Delta x_j}
\end{equation}
First, we look at the on-diagonal elements of this matrix, which are the temporal variances of different neurons.
To determine the average temporal variance, we take the sample mean across neurons and we average over the random connectivity, obtaining (see Eq.(\ref{ondiag2A}) in Appendix 1)

\begin{equation}
\label{ondiag2}
\left<\overline{\Delta x^2}\right>=\left<\frac{1}{N}\sum_{i=1}^NQ_{ii}\right>=\frac{\overline{\Delta x_{ext}^2}}{2}\left[\frac{k_{ext}g_{ext}^2}{1+g\sqrt{K}}\;\xi+\frac{\lambda_{ext}^2}{\sqrt{1-\lambda^2}}\right]
\end{equation}
The temporal variance of neural activity is the sum of two pieces: the first term decreases with the number of connections as $K^{-1/2}$, while the second term remains finite (the factor $\xi$ is close to one, see Eq.(\ref{xi}) in Appendix 1).
The first term indicates that recurrent inhibition ($g$) reduces temporal fluctuations.
In fact, the inhibitory feedback not only reduces the mean activity (Eq.(\ref{mx})), but also cuts down fluctuations by quickly counterbalancing the external excitatory input.
This can be verified by calculating the instantaneous covariance between the external excitatory and the local inhibitory input, which is found large and negative, equal to $-k_{ext}g_{ext}^2g\sqrt{K}$.
The second term implies that non-zero fluctuations arise even in large networks (large $K$), and inhibition cannot exert an instantaneous and exact balance for each neuron. 
However, fluctuations nearly vanish if the external input is homogeneous ($\lambda_{ext}=0$), in which case the inhibitory feedback would definitively counterbalance the homogeneous drive.
Furthermore, as in the case of spatial fluctuations, temporal fluctuations increase with the heterogeneity of connections, recurrent ($\lambda$) and external ($\lambda_{ext}$).
Temporal fluctuations diverge when the network approaches the instability point, when the linear approximation fails ($\lambda\rightarrow 1$).

\begin{figure}[h!]
\begin{center}
\includegraphics[width=4in]{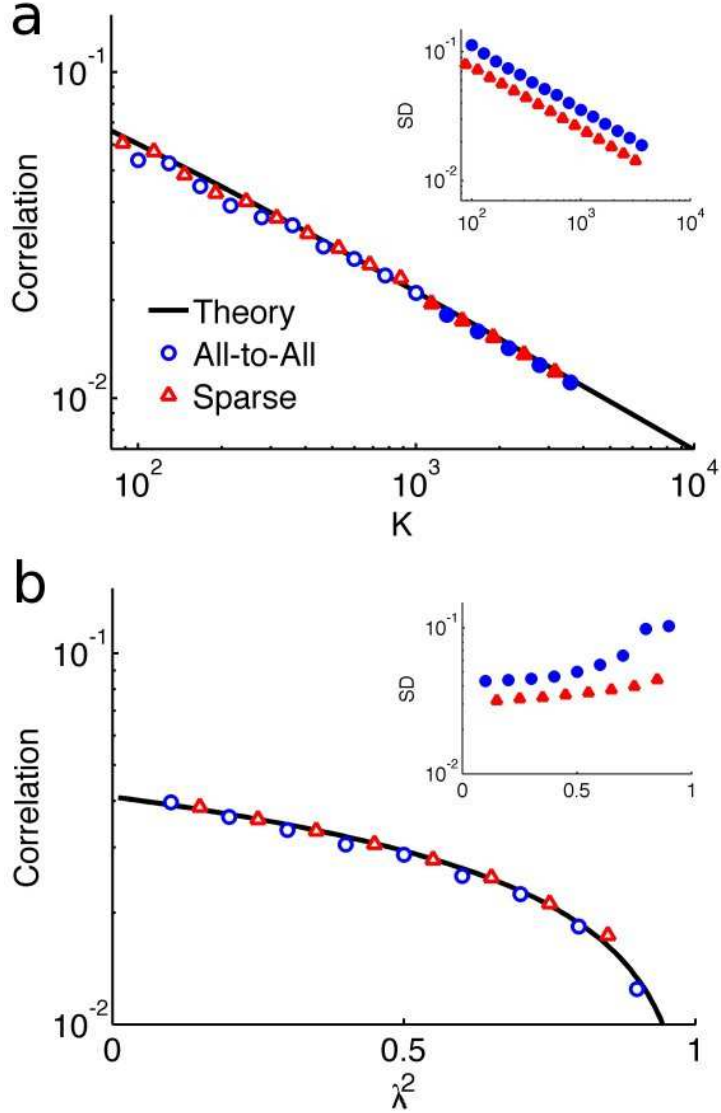}
\end{center}
\caption{
The mean correlation in the activity between neuron pairs decreases with the number of connections $K$ (a) and their heterogeneity $\lambda$ (b), Standard Deviation (SD) is shown in the insets.
Analytical results (lines) are obtained from Eq.(\ref{corr}).
Simulation results are shown for the All-to-All (blue circles) and Sparse (red triangles) network.
Open symbols show simulations of the neural dynamics, Eq.(\ref{xdyn1});
Filled symbols show numerical evaluation of Eq.(\ref{C_OU}).
The parameters used are, in both panels, $g=g_{ext}=1$, $\overline{x}_{ext}=\overline{\Delta x_{ext}^2}=1$.
In panel b, $K_{ext}=K=500$.
For the All-to-All network: $\lambda_{ext}=1$, and $\lambda=1/\sqrt{2}$ in panel a.
For the Sparse network: $k_{ext}=k=1/2$ in panel a, which correspond to $\lambda=\lambda_{ext}=1/\sqrt{2}$, while $k_{ext}=k$ is varied in panel b according to the value of $\lambda$ (see Methods).}
\label{Fig_cor}
\end{figure}

How much of the total variance, expressed in Eq.(\ref{ondiag2}), is independent rather than shared between neurons? 
To answer this question we calculate the average covariance, by looking at the off-diagonal elements of the covariance matrix, the pairwise covariances.
We take the sample mean across neuron pairs and average over the random connectivity to obtain the average covariance (see Eq.(\ref{offdiagA}) in Appendix 1)

\begin{equation}
\label{offdiag}
\left<\overline{\Delta x'\Delta x''}\right>=\left<\frac{1}{N(N-1)}\sum_{i\neq j}^{1,N}Q_{ij}\right>=\frac{\overline{\Delta x_{ext}^2}}{2}\;\frac{k_{ext}g_{ext}^2}{1+g\sqrt{K}}
\end{equation}
Notably, this is proportional to the first term in the total variance, Eq.(\ref{ondiag2}), by a factor close to one ($\xi\simeq 1$, see Eq.(\ref{xi}) in Appendix 1), implying that the two terms in the total variance express, respectively, the correlated and uncorrelated fluctuations. 
Therefore, while the uncorrelated variance remains finite for large $K$, the correlated variance vanishes.
The activities of neuron pairs tend to covary, due to their shared external input, but the recurrent inhibition makes the covariance small, of order $K^{-1/2}$.
In the Sparse network, the covariance vanishes if the probability of external connections is small ($k_{ext}\rightarrow0$), since the shared external input between neurons tends to zero in that case.
In the All-to-All network, the mean covariance vanishes if the mean input connection is zero ($g_{ext}=0$);
In that case, even if neurons receive a shared external input, neuron pairs may weight different inputs with the same or opposite signs, leading to respectively positive or negative covariance.
Therefore, while the mean covariance across neuron pairs is zero, the covariance of single pairs may be positive or negative.

The mean correlation is obtained by dividing the covariance, Eq.(\ref{offdiag}), by the variance, Eq.(\ref{ondiag2}), i.e. (we assume that variance and covariance are independent): 

\begin{equation}
\label{corr}
\left<R\right>=\frac{\left<\overline{\Delta x'\Delta x''}\right>}{\left<\overline{\Delta x^2}\right>}=\frac{1}{\xi+\frac{\lambda_{ext}^2}{\sqrt{1-\lambda^2}}\frac{(1+g\sqrt{K})}{k_{ext}g_{ext}^2}}
\end{equation}
This expression is positive and never exceeds one.
It indicates that the mean correlation is small, of order $K^{-1/2}$, despite the strong and shared excitatory input between neurons.
However, this result holds only in presence of the local recurrent inhibition ($g>0$), and provided that external connections are heterogeneous ($\lambda_{ext}^2\neq0$).
Heterogeneity of local connections ($\lambda$) also contributes in decreasing the correlation.

Therefore, the inhibitory feedback and the random connectivity are responsible for the small correlation.
If the inhibitory feedback is removed, $g=0$, the correlation becomes large.
If the network heterogeneity is removed, $\lambda=\lambda_{ext}=0$, the correlation is equal to one, because the network is homogeneous and all neurons get the same input ($\xi=1$ when $\lambda=0$, see Eq.(\ref{xi}) in Appendix 1).
Fig.\ref{Fig_cor} shows an example of the mean correlation as a function of the number of connections and the heterogeneity of the network.
The analytical result, Eq.(\ref{corr}), agrees with numerical simulations of the linear dynamics, in both the All-to-All and the Sparse network.
Insets in Fig.\ref{Fig_cor} show the Standard Deviation of correlations, which appear to decrease with the number of connections as $K^{-1/2}$, and to increase with the heterogeneity of the network.

\begin{figure}[h!]
\begin{center}
\includegraphics[width=5.5in]{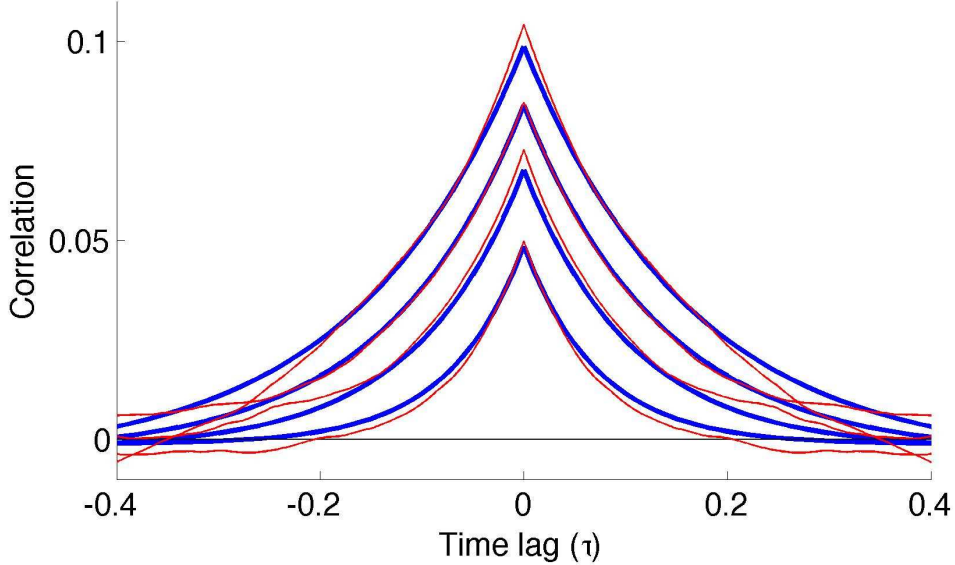}
\end{center}
\caption{Both peak and width of the mean cross-correlation decrease with the number of connections $K$.
The plot shows the cross-correlation as a function of time lag;
Different curves, from top to bottom, correspond to $K=100, 156, 278, 625$ in the All-to-All network.
Consistent with Fig.\ref{Fig_cor}, the correlation at zero lag decreases with $K$.
This plot shows that also the temporal width of the cross correlation decreases with $K$.
Red lines: Simulations of neural dynamics;
Blue lines: Numerical evaluation of Eq.(\ref{C_lag}).
Parameters: $g=0.5$, $g_{ext}=1$, $\overline{x}_{ext}=\overline{\Delta x_{ext}^2}=1$, $\lambda_{ext}=1$, $\lambda=1/\sqrt{2}$.
}
\label{crosscor}
\end{figure}

The final issue that we address is the timescale of correlations.
Neural activity integrates the input on multiple timescales, because of the large number of neurons and the heterogeneity of their connections.
Which timescales are responsible for correlations? What correlations characterize the activity integrated in time? 
Note that the mean correlation in Eq.(\ref{corr}) and Fig.\ref{Fig_cor} is the correlation at zero lag, namely the instantaneous correlation.
We investigate the timescale of correlations in Fig.\ref{crosscor}, where the cross-correlation of neural activity is shown at different time lags.
The correlation has a peak at zero lag, and shows an exponential decay in time.
As we have shown in Fig.\ref{Fig_cor}, the correlation at zero lag decreases with the number of connections as $K^{-1/2}$.
Fig.\ref{crosscor} shows that the timescale of correlation, determining its rate of decay, also decreases with the number of connections.
In fact, we show in Appendix 1 that the integrated correlation across all time lags, namely the total area of the cross-correlation, is of magnitude $K^{-1}$.
Since the total area is approximately equal to correlation peak times temporal width, then the temporal width is of order $K^{-1/2}$.
Therefore, inhibition decorrelates on a fast timescale, and integrating neural activity, even for a relatively short time, has the effect of further decreasing the magnitude of correlations (see Appendix, Eq.(\ref{corrY})).

\begin{figure}[h!]
\begin{center}
\includegraphics[width=5.5in]{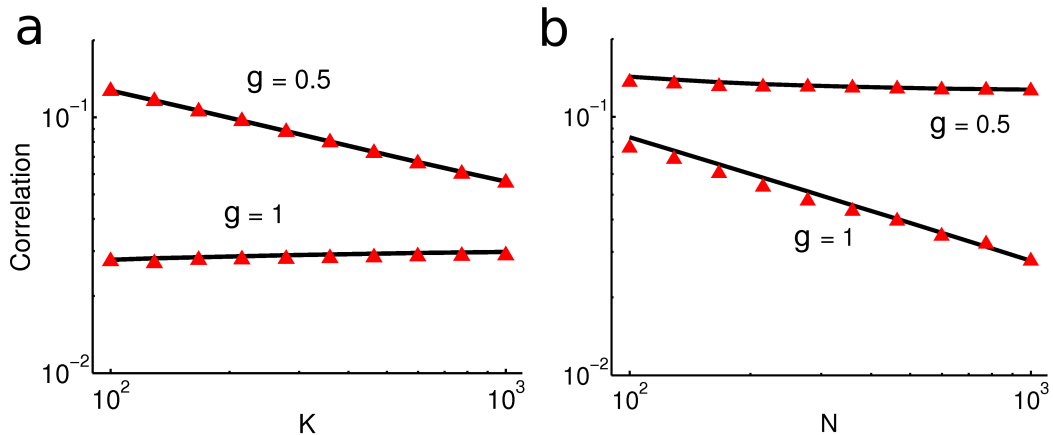}
\end{center}
\caption{The mean correlation depends primarily on the number of connections $K$ (a) or the number of neurons $N$ (b) depending on the strength of inhibition.
If inhibition is weak ($g=0.5$), correlations depend mostly on $K$, while if inhibition is stronger ($g=1$), correlations depend mostly on $N$. 
Analytical results (lines) are obtained from Eq.(\ref{corr}).
Filled symbols show numerical evaluation of Eq.(\ref{C_OU}) for the Sparse network.
Panel a: $N=1000$.
Panel b: $K=100$.
Other parameter values are: $g_{ext}=g=$value in figure, $\overline{x}_{ext}=\overline{\Delta x_{ext}^2}=1$, $N_{ext}=1000$, $K_{ext}=500$, $k_{ext}=0.5$.
The values of $\lambda$ and $\lambda_{ext}$ vary according to the values of $k$, $\rho$ and $\rho_{ext}$ (see Methods).
}
\label{corKN}
\end{figure}

It is worth noting that, while in other studies the results are often described in terms of the number of neurons $N$, here both $N$ and the number of connections $K$ play a role.
For the Sparse network, it is interesting to note that all the above results depend on the number of neurons $N$ only through the parameter $\lambda^2=g^2(1-K/N)$, because $N$ affects the sparsity of connections and therefore also their variance.
The order of magnitude of correlations $K^{-\frac{1}{2}}$ holds regardless of the number of neurons, which may be taken even infinite for any fixed value of $K$.
However, the dependence of correlations on the heterogeneity $\lambda$, and therefore $N$, may be quite substantial.
Fig.\ref{corKN} shows how the mean correlation varies as a function of either $K$ or $N$:
for a relatively weak inhibition, the mean correlation depends mostly on the number of connections $K$, while for stronger inhibition the mean correlation depends mostly on the number of neurons $N$.

\subsection*{Spiking network simulations}

\begin{figure}[h!]
\begin{center}
\includegraphics[width=6in]{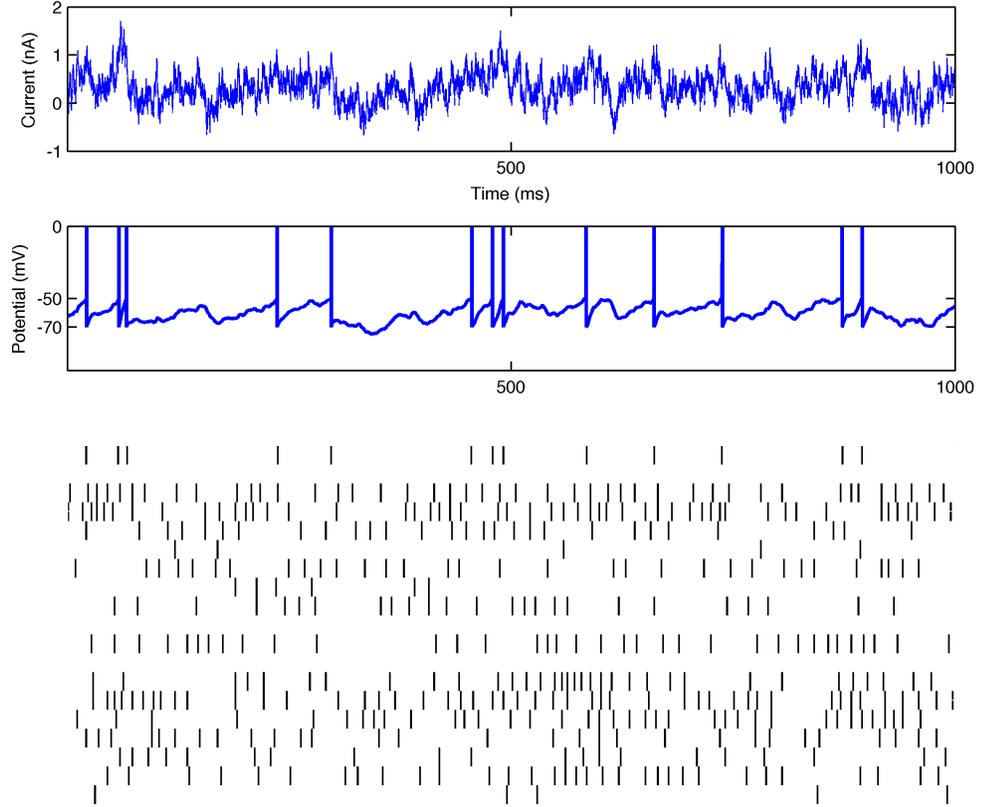}
\end{center}
\caption{
Example of dynamics in the spiking network simulation.
Top: Dynamics of the total input current (excitatory and inhibitory) to one neuron, in a time span of one second.
Middle: Dynamics of the membrane potential in the same neuron and temporal window. 
In the integrate and fire model, action potentials are instantaneous and of arbitrary size.
Bottom: spike times (rasters) of twenty example neurons, each row represents one neuron and each tick represents one action potential for the corresponding neuron.
The neuron in the top row corresponds to the neuron depicted in the top and middle part of the figure.
Parameters used in simulations are: $\tau=10ms$, $V_{th}=-50mV$, $V_{rs}=-70mV$, $V_L=-70mV$, $\phi_{ext}=50Hz$, $C=0.4nF$, $g=20nS$.
Synaptic parameters are: $g=1$, $\lambda=0.5$, $g_{ext}=1$, $\lambda_{ext}=0.58$ (in units of $8\;nA\cdot ms$).}
\label{spiking1}
\end{figure}

We tested the predictions of the linear model in a more realistic spiking network, described by a current-based Integrate and Fire model (see Methods).
The spiking network is characterized by the non-linear dynamics inherent in the generation of action potentials.
However, we tested the hypothesis that this non-linear system, when displaying small fluctuations around a steady state, may be approximated by a linear system and therefore by the equations derived in the previous section.
Fig.\ref{spiking1} shows the dynamics of an example neuron's input current and membrane potential, and spike times (rasters) of that neuron and other neurons from the spiking network.
Simulation results reproduce qualitatively the phenomenology observed in the Cerebral Cortex:
Since the input current puts neurons close to the firing threshold, neurons are susceptible to noise and fire irregularly, with noisy spike emission times \cite{shadlen98};
The distribution of firing rates across neurons is broad, with a higher proportion of neurons displaying low firing rates \cite{baddeley97,hromadka08}.

Our goal is not to provide a formal theory for the linear approximation of a spiking model.
Instead, we show the results of a quantitative comparison of the two models, and we briefly summarize below the theoretical arguments underlying this comparison.
Since all results of the linear model are stated in terms of the mean and variance of the synaptic matrix, we hypothesize that the linear response of the spiking network can be described by an "effective" set of synaptic parameters $g, \lambda, g_{ext}, \lambda_{ext}$. 
For a given network size ($N=200$), we probed the response of the spiking network to small changes in the external input, and we used these linear responses to fit the effective parameters of the connectivity.
Then, we used these parameters to predict the mean correlation across a wide range of network sizes (from $N=50$ to $N=1000$), without fitting any additional parameter.

\begin{figure}[h!]
\begin{center}
\includegraphics[width=6in]{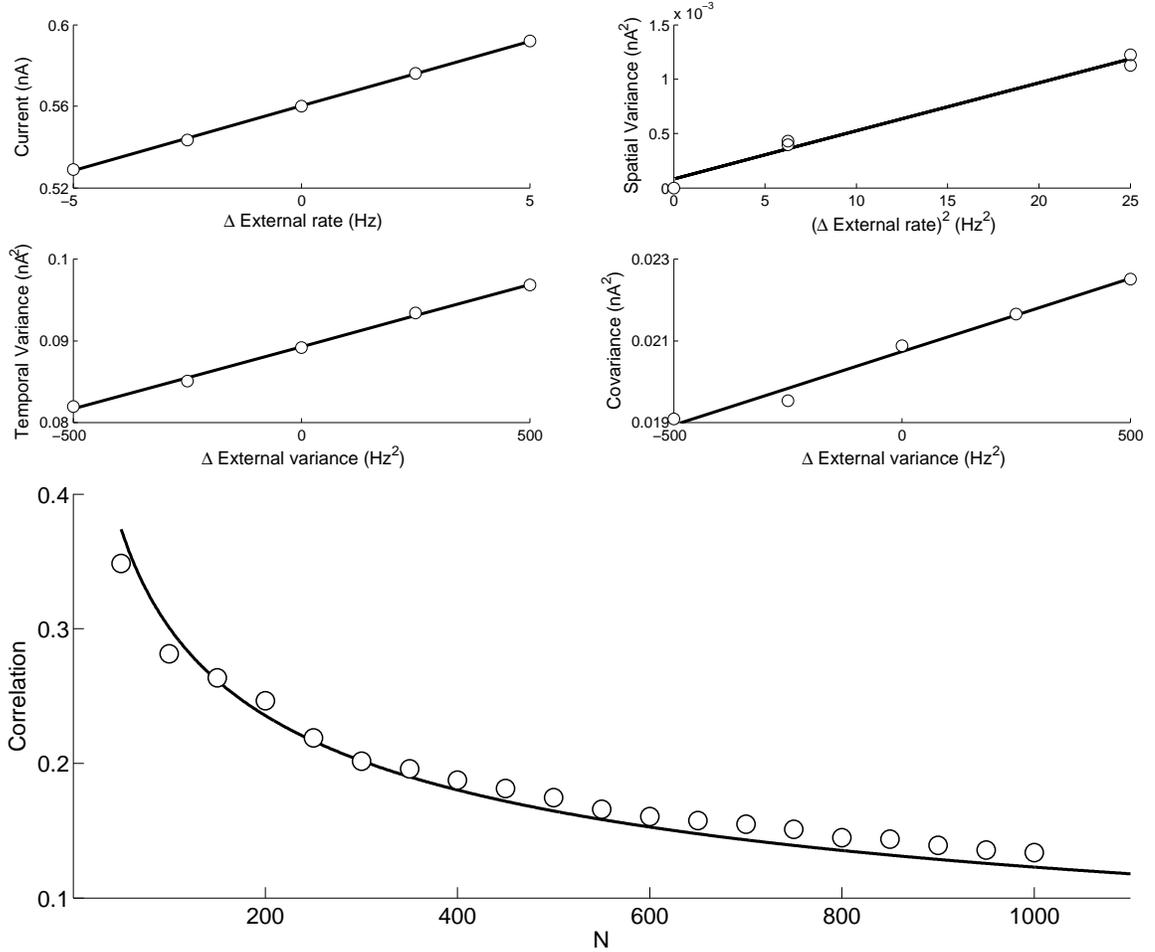}
\end{center}
\caption{Linear response of the spiking network to external input at fixed $N=200$ (a-d) and correlations predicted from the linear response at variable network sizes (e) from $N=50$ to $1000$. Four statistics of the current are presented: mean current (a), spatial variance (b), temporal variance (c) and covariance (d). Linear regression is used in (a-d) 
to fit effective parameters and predict correlations at variable network sizes in (e). Circles: spiking network simulations; Lines: Linear regression fit (a-d) and analytical prediction of correlation (e).
}
\label{spiking2}
\end{figure}

Fig.\ref{spiking2} (panels a-d) shows the change in the total current (excitatory and inhibitory) as a consequence of the change in the external input rate.
This change is approximately linear; 
Panels a-d show four statistics of the current, respectively, the change in mean current (a), spatial variance (b), temporal variance (c) and covariance (d).
In the linear model, those quantities are calculated, respectively, in Eqs.(\ref{mx}),(\ref{s2x}),(\ref{ondiag2}),(\ref{offdiag}), where $x_{ext}$ corresponds to the change in external input rate.
We use linear regression to fit the slopes in Fig.\ref{spiking2}a-d, and invert the equations of the linear model to obtain the effective values of the parameters of the connectivity (see Methods).
The fitted values are: $g=2.44$, $\lambda=0.54$, $g_{ext}=1.99$, $\lambda_{ext}=0.55$, to be compared with those used to generate the synaptic matrices in the spiking model (see Methods, $g=1$, $\lambda=0.5$, $g_{ext}=1$, $\lambda_{ext}=0.58$ in units of $8\;nA\cdot ms$).
Then, we use these values to predict the mean correlation across a wide range of network sizes, by using the formula for the linear model Eq.(\ref{corr}).
The result is plotted in Fig.\ref{spiking2}e, showing a remarkable agreement with theory.
The fact that the theory provides a good fit for $N=200$ is obvious, since parameters are fit at that specific network size (although in a different simulation).
However, the good fit across a wide range of network sizes suggests that equations of the linear model provide a good instrument to probe spiking networks.
 
The theoretical arguments underlying the above analysis are based on the dynamics of the total current integrated by the membrane potential of a neuron, both the inhibitory recurrent and the excitatory external current (see e.g. Fig.\ref{spiking1} top).
This dynamics is described by Eq.(\ref{Idyn}) in Methods, which is similar to the equation describing the linear system (Eq.(\ref{xdyn1})):
In the spiking network, the external stimulus  is characterized by a sum over $N_{ext}$ independent Poisson spike trains weighted by the matrix $G^{ext}$;
If $N_{ext}$ is large enough, this is approximately equal to a Gaussian white noise process \cite{amit91}, and therefore is equivalent to the external input of the linear model. 
The main difference between the linear and spiking model is the input from neurons in the recurrent network, determined by the spike trains $S_i(t)$.
Those spike trains are non-linear and non-instantaneous functions of the input currents, and also provide an additional source of noise, due to the discrete spike times.
Nevertheless, we found that the parameters of an effective linear system, determined by the linear response of the spiking network, are able to predict well the correlations.

\section*{Discussion}
We found that inhibitory feedback and heterogeneous connections have important effects on the dynamics of the activity in a neural circuit.
The strong excitatory input, shared between neurons, tends to drive the network to a highly active and correlated state.
The inhibitory feedback is responsible for balancing the network activity, and also for reducing temporal fluctuations, in particular the correlated fluctuations across neurons.
The heterogeneity of couplings plays a crucial role in reducing correlations, since homogeneous connections would determine homogeneous and therefore highly correlated activity.
As a consequence, the observed mean correlation is positive and of small magnitude.
The fact that mean correlation is positive is obvious, since neurons in a large population cannot be anti-correlated on average\footnote{The mean correlation must be larger than $-\frac{1}{N-1}$, therefore it must be non-negative in infinitely large populations. Proof: Any covariance matrix $Q$ is positive definite, therefore $h^\dagger Qh>0$ for any vector $h$. If we choose $h_i=1/\sqrt{Q_{ii}}$ and define the correlation matrix $R_{ij}=Q_{ij}/\sqrt{Q_{ii}Q_{jj}}$ we have that $\sum_{ij}R_{ij}=N+N(N-1)\left<R\right>>0$. Therefore the mean correlation must be $\left<R\right>>-\frac{1}{N-1}$. Tighter bounds on the mean correlation can be obtained by using $\lambda_{min}\leq h^\dagger Qh/|h|^2\leq\lambda_{max}$, where $\lambda_{min}$ and $\lambda_{max}$ are, respectively, the minimum and maximum eigenvalue of $Q$. This implies $(\lambda_{min}/\lambda_{max}-1)/(N-1)\leq\left<R\right>\leq(\lambda_{max}/\lambda_{min}-1)/(N-1)$.}.
What is not obvious is that the mean correlation is of small magnitude.

The main contribution of our work is an analytical calculation of the effect of heterogeneity on correlations, in terms of random connectivity or random synaptic strengths. 
In presence of heterogeneous connections and inhibitory feedback, the mean correlation at zero time lag is small, it decreases with the number of connections as $K^{-\frac{1}{2}}$.
The mean correlation integrated on large timescales is even smaller, of order $K^{-1}$, indicating that inhibition downsizes correlations on a timescale of $K^{-\frac{1}{2}}$.
Other modeling studies have addressed the issue of correlations in neural circuits. 
In previous studies \cite{ginzburg94,hertz10,renart10,tetzlaff12}, the mean correlation was found to decrease with the number of neurons as $N^{-1}$.
In \cite{ginzburg94, hertz10}, a network is studied in which connections strengths are of order $N^{-1}$, which implies a weak interaction between neurons and therefore a weak correlation. 
More surprisingly, \cite{renart10, tetzlaff12} found weak correlations even in the case of strong interactions, with connection strengths of order $N^{-1/2}$.
Both studies found a mean correlation of magnitude $N^{-1}$, provided that the correlation is integrated across large temporal windows.
However, \cite{renart10} shows that mean correlation at zero time lag, of membrane currents, is of magnitude $N^{-1/2}$.
These results are consistent with our findings, compare for example Fig.\ref{crosscor} with Fig.2E of \cite{renart10} and Fig.3D of \cite{tetzlaff12}.  
However, we highlight a potential difference by noting that the main parameter affecting correlations may be the number of connections $K$ rather than the number of neurons $N$.
The exclusive contribution of these two parameters has not been studied in detail in previous studies, and we have shown that it may depend on the network regime.

We also studied a more realistic spiking network model, and we confronted the analytical solutions of the linear model with the simulations of the non-linear spiking model.
We looked at small, linear changes in the current as a function of changes in the external firing rate input to the spiking network, and computed the effective parameters of the linear model able to explain those changes.
We found that those effective, "linearized" parameters are able to predict correlations accurately even when changing the network size significantly.
This suggests that the linear approximation is adequate for studying correlations.
We did not consider the problem of a complete linear theory of spiking models, which would address the issues of computing the linearized kernel and the effective interaction matrix.
Those issue have been studied for example in \cite{lindner05, trousdale12, tetzlaff12}.

The small mean correlation observed in our study and previous modeling studies agrees qualitatively with the experimental observations.
The model may be useful to investigate correlations in different brain areas, especially those dominated by inhibition, such as the Striatum and Globus Pallidus.
Interestingly, large correlations between pallidal neurons have been shown to be correlated with Parkinsonism \cite{raz00, wilson11}.
The model is also consistent with the strong anti-correlations between excitatory and inhibitory inputs observed experimentally \cite{okun08, cafaro10, salinas00}.
However, different experimental studies report quantitative differences in measured correlations.
For example, correlations depend on the temporal window on which action potentials are counted to determine a neuron's firing rate (see Fig.\ref{eightstudies}).
While neural dynamics occurs on a variety of timescales in our model\footnote{Since the dynamics is linear, see Eq.(\ref{xdyn2}), the timescales of the network are determined by the eigenvalues of the matrix $(I-G)$. 
A random matrix with Gaussian and independent elements has eigenvalues distributed uniformly in a circle in the complex plane centered at $0$ and of radius $\lambda$ (where its elements have variance $\lambda^2/N$, see \cite{novak03}). 
One isolated eigenvalue is found approximately at $mN$, where $m$ is the mean of the elements of the matrix and $N$ is its dimension. 
Finally, the identity matrix translates all eigenvalues by one.
Therefore, the eigenvalues of $(I-G)$ are contained in a circle centered at $1$ and of radius $\lambda$, with one additional eigenvalue approximately equal to $1+g\sqrt{K}$.
Timescales, in units of $\tau$, are equal to the inverse of those eigenvalues, therefore the fastest timescale is equal to $1/(1+g\sqrt{K})$, while the remaining timescales are in between $(1+\lambda)^{-1}$ and $(1-\lambda)^{-1}$.
}
, as well as in real neurons \cite{bernacchia11}, additional modeling studies are necessary to capture the wide range of phenomena observed in the experimental measures of correlations, including the effects of distance between neurons, multiple timescales and firing activity \cite{cohen11}.

\section*{Acknowledgements}
This study was supported by the US National Institutes of Health grant R01 MH062349 and the Swartz Foundation.

\section*{Appendix 1: Statistics of random networks}
In this section, we calculate the averages of neural activity and correlations with respect to both temporal fluctuations (noise) and the spatial variability of the connection strengths (disorder).
Due to the linearity of the model, all quantities of interests can be simply calculated;
The novel contribution of this work is averaging those quantities over the randomness of the connectivity matrix.
The equation of dynamics (\ref{xdyn1}) can be expressed in matrix form (the time constant of temporal evolution $\tau$ is set to $1$):

\begin{equation}
\label{xdyn2}
\frac{d\textbf{x}(t)}{dt}=(G-I)\textbf{x}(t)+G_{ext}\textbf{x}_{ext}(t)
\end{equation}
where $\textbf{x}$ is the vector of local neural activities, $\textbf{x}_{ext}$ is the vector of external neural activities, the matrices $G$ (of size $N\times N$) and $G_{ext}$ (size $N\times N_{ext}$) express respectively the recurrent connections and the feed-forward projections, and $I$ is the identity matrix.
The equation of dynamics is linear and, given the interaction matrices $G$, $G_{ext}$ and the input signal $\textbf{x}_{ext}$, the neural activity can be expressed as 
a sum over the external input weighted by an exponential temporal decay

\begin{equation}
\label{xdynsol}
\textbf{x}(t)=\int_{-\infty}^tdt'e^{(G-I)(t-t')}G_{ext}\textbf{x}_{ext}(t')=\int_0^{+\infty}dt' e^{(G-I)t'}G_{ext}\textbf{x}_{ext}(t-t')
\end{equation}
We assumed that initial conditions have decayed and that the inequality $\lambda<1$ holds, to prevent network activity from growing in time without bounds.
In the limit of large $N$, the real part of the eigenvalues of $G$ is bounded by $\lambda$ \cite{novak03}.
Therefore, if $\lambda\geq1$, some eigenvalues of $(G-I)$ have non-negative real part, and the integral does not converge.
This corresponds to an unstable fixed point at $\textbf{x}=0$ and network activity grows in time without bounds.

We start by calculating the mean neural activity.
We perform the temporal average of the above expression, therefore we substitute the external activity $\textbf{x}_{ext}(t)$ with its average $\overline{x}_{ext}$, and we perform the integral, obtaining (temporal average is denoted by overline)

\begin{equation}
\label{thermalx}
\overline{\textbf{x}}=\overline{x}_{ext}\;(I-G)^{-1}G_{ext}\mathbf{1}
\end{equation}
where the vector $\mathbf{1}$ has all $N_{ext}$ components equal to one.
Because the matrices of connection strengths are heterogeneous, $G$ and $G_{ext}$, different neurons have a different mean activity.
In order to calculate the spatially averaged activity, we compute the sample mean across neurons.
For large $N$, this is independent on the specific realization of the spatial disorder, therefore we perform its average over the distribution of connectivity strengths, namely

\begin{equation}
\left<\overline{x}\right>=\left<\frac{1}{N}\sum_{i=1}^N\overline{x_i}\right>=\left<\frac{\overline{x}_{ext}}{N}\mathbf{1}^\dagger(I-G)^{-1}G_{ext}\mathbf{1}\right>
\end{equation}
The average (angular brackets) is across all possible realizations of the random matrices $G$ and $G_{ext}$.
We denote by $\dagger$ the transpose operation.
Note that, in the expression above, the row vector $\mathbf{1}^\dagger$ has $N$ components, while the column vector $\mathbf{1}$ has $N_{ext}$.
In the following we will use the same notation regardless of the dimension of $\mathbf{1}$, since that can be determined by the dimension of the multiplied
 matrix.
Since $G$ and $G_{ext}$ are independent, we can substitute $G_{ext}$ with its mean, $\left<G_{ext}\right>=\frac{g_{ext}\sqrt{K_{ext}}}{N_{ext}}\mathbf{1}\mathbf{1}^\dagger$.
Furthermore, we show in Appendix 2, Eq.(\ref{appeq1}), that $<\mathbf{1}^\dagger(I-G)^{-1}\mathbf{1}>=N(1+g\sqrt{K})^{-1}$.
Therefore, the mean activity is equal to

\begin{equation}
\label{mxA}
\left<\overline{x}\right>=\frac{g_{ext}\sqrt{K_{ext}}}{1+g \sqrt{K}}\,\overline{x}_{ext}
\end{equation}
This expression is used in the main text (Eq.(\ref{mx})).

Different neurons have different connections and therefore different activity, and the extent to which the activity varies from neuron to neuron is determined by the spatial variance.
We calculate this quantity by taking the sample variance across neurons and averaging over the spatial disorder.
We take the scalar product of Eq.(\ref{thermalx}) with itself, and we use again the fact that the sample mean does not depend on the spatial disorder for large $N$, to obtain

\begin{equation}
\left<\Delta\overline{x}^2\right>=\left<\frac{\textbf{x}^\dagger \textbf{x}}{N}\right>-\left<\overline{x}\right>^2=\left<\frac{\overline{x}_{ext}^2}{N}\mathbf{1}^\dagger G_{ext}^\dagger(I-G^\dagger)^{-1}(I-G)^{-1}G_{ext}\mathbf{1}\right>-\left<\overline{x}\right>^2
\end{equation}
We rewrite this expression by using the trace operator and its cyclic invariance.
Namely, for any arbitrary matrices $A$, $B$, the following equations hold: $\mathbf{1}^\dagger A\mathbf{1}=\mbox{Tr}(A\mathbf{1}\mathbf{1}^\dagger)$ and $\mbox{Tr}(AB)=\mbox{Tr}(BA)$.
We obtain

\begin{equation}
\left<\Delta\overline{x}^2\right>=\left<\frac{\overline{x}_{ext}^2}{N}\mbox{Tr}\left((I-G^\dagger)^{-1}(I-G)^{-1}G_{ext}\mathbf{1}\mathbf{1}^\dagger G_{ext}^\dagger\right)\right>-\left<\overline{x}\right>^2
\end{equation}
Again, since $G$ and $G_{ext}$ are independent, we can average separately the factors involving the two matrices.
A simple calculation shows that $\left<G_{ext}\mathbf{1}\mathbf{1}^\dagger G_{ext}^\dagger\right>=g_{ext}^2K_{ext}\mathbf{1}\mathbf{1}^\dagger+\lambda_{ext}^2I$.
Furthermore, we show in Appendix 2, Eqs.(\ref{appeq2},\ref{appeq3}) that the following two equalities hold $<\mbox{Tr}((I-G)^{-1}(I-G^\dagger)^{-1})>=N(1-\lambda^2)^{-1}$, and $<\mbox{Tr}((I-G^\dagger)^{-1}(I-G)^{-1}\mathbf{1}\mathbf{1}^\dagger)>=N(1-\lambda^2)^{-1}(1+g\sqrt{K})^{-2}$.
Using the expression of the mean activity, Eq.(\ref{mxA}), the spatial variance is equal to

\begin{equation}
\label{s2xA}
\left<\Delta\overline{x}^2\right>=\frac{1}{1-\lambda^2}\left[\left<\overline{x}\right>^2\lambda^2+\overline{x}_{ext}^2\lambda_{ext}^2\right]
\end{equation}
This expression is used in the main text (Eq.(\ref{s2x})).

After looking at the spatial variability, we study temporal variability and correlated fluctuations by calculating the covariance matrix.
We take the scalar product of Eq.(\ref{xdynsol}) with itself and we perform the temporal average, using the fact that the external stimulus is uncorrelated in space and time.
This corresponds to the covariance matrix of a Ornstein-Uhlenbeck process \cite{gardiner}, and is equal to

\begin{equation}\label{C_OU}
Q=\overline{\Delta\textbf{x}\Delta\textbf{x}^\dagger}=\overline{\Delta x_{ext}^2}\int_0^{+\infty}dt\; e^{(G-I)t}G_{ext}G_{ext}^\dagger\; e^{(G^\dagger-I)t}\end{equation}
Note that the covariance matrix satisfies the Lyapunov equation

\begin{equation}\label{lyapunov}
(G-I)Q+Q(G^\dagger-I)+\overline{\Delta x_{ext}^2}G_{ext}G_{ext}^\dagger=0
\end{equation}
but this cannot be used for averaging $Q$, since $G$ and $Q$ are dependent and they do not commute.
Note also that Eq.(\ref{C_OU}) represents the covariance at zero time lag; We will consider the case of finite time lag at the bottom of this section.

The on-diagonal elements of the covariance matrix are the temporal variances of different neurons.
To determine the average temporal variance, we take the sample mean across neurons and we average over the spatial disorder, obtaining
\begin{equation}
\left<\overline{\Delta x^2}\right>=\left<\frac{1}{N}\mbox{Tr}(Q)\right>=\frac{\overline{\Delta x_{ext}^2}}{N}\int_0^{+\infty}dt \left<\mbox{Tr}\left(e^{(G^\dagger-I)t}e^{(G-I)t}G_{ext}G_{ext}^\dagger\right)\right>
\end{equation}
where we applied the trace operator to select the diagonal elements, and we used its cyclic invariance.
Again, since $G$ and $G_{ext}$ are independent, we can average separately the factors involving the two matrices.
A simple calculation gives $\left<G_{ext}G_{ext}^\dagger\right>=k_{ext}g_{ext}^2\mathbf{1}\mathbf{1}^\dagger+\lambda_{ext}^2I$.
Furthermore, using Eqs(\ref{appeq4},\ref{appeq5}) in Appendix 2, we obtain

\begin{equation}
\label{ondiag2A}
\left<\overline{\Delta x^2}\right>=\frac{\overline{\Delta x_{ext}^2}}{2}\left[\frac{k_{ext}g_{ext}^2}{1+g\sqrt{K}}\;\xi+\frac{\lambda_{ext}^2}{\sqrt{1-\lambda^2}}\right]
\end{equation}
This corresponds to Eq.(\ref{ondiag2}) in the main text.
The factor $\xi$ is equal to

\begin{equation}\label{xi}
\xi=\left[1-\frac{\lambda^2}{1+\sqrt{1-\lambda^2}(1+g\sqrt{K})}\right]^{-1}
\end{equation}
It is never smaller than one, and it is very close to one for a wide range of parameters, including for large $K$ and for small $\lambda$.
However, it diverges near the critical point $\lambda\simeq 1$.

Next, we calculate the average covariance, by looking at the off-diagonal elements of the matrix in Eq.(\ref{C_OU}).
The off-diagonal elements are the pairwise covariances, and the sample mean across neuron pairs can be averaged over the spatial disorder to obtain the average covariance.
We use the matrix $(\mathbf{1}\mathbf{1}^\dagger-I)$ to select the off-diagonal elements, and we obtain

\begin{eqnarray*}\left<\overline{\Delta x'\Delta x''}\right>&=&\left<\frac{1}{N(N-1)}\mbox{Tr}\left((\mathbf{1}\mathbf{1}^\dagger-I)Q\right)\right>=
\\&=&\frac{\overline{\Delta x_{ext}^2}}{N(N-1)}\int_0^{+\infty}dt \left<\mbox{Tr}\left(e^{(G^\dagger-I)t}(\mathbf{1}\mathbf{1}^\dagger-I)e^{(G-I)t}G_{ext}G_{ext}^\dagger\right)\right>\end{eqnarray*}
Again, we used the cyclic invariance of the trace operator and, since $G$ and $G_{ext}$ are independent, we can average separately the factors involving the two matrices.
We use $\left<G_{ext}G_{ext}^\dagger\right>=k_{ext}g_{ext}^2\mathbf{1}\mathbf{1}^\dagger+\lambda_{ext}^2I$.
Furthermore, using Eqs.(\ref{appeq4},\ref{appeq5},\ref{appeq6}) in Appendix 2 and neglecting all terms of order $N^{-1}$, we obtain

\begin{equation}
\label{offdiagA}
\left<\overline{\Delta x'\Delta x''}\right>=\frac{\overline{\Delta x_{ext}^2}}{2}\;\frac{k_{ext}g_{ext}^2}{1+g\sqrt{K}}
\end{equation}
This expression is used in the main text (Eq.(\ref{offdiag})).

The mean correlation is obtained by dividing the covariance, Eq.(\ref{offdiagA}), by the variance, Eq.(\ref{ondiag2A}), i.e. (we assume that variance and covariance are independent): 

\begin{equation}
\label{corrA}
\left<R\right>=\frac{\left<\overline{\Delta x'\Delta x''}\right>}{\left<\overline{\Delta x^2}\right>}=\frac{1}{\xi+\frac{\lambda_{ext}^2}{\sqrt{1-\lambda^2}}\frac{(1+g\sqrt{K})}{k_{ext}g_{ext}^2}}
\end{equation}
This expression is positive and never exceeds one.
This corresponds to Eq.(\ref{corr}) in the main text.

We now turn to calculating the correlations of activity integrated in time.
In order to calculate those correlations, we define the covariance at time lag $\Delta t$ as $Q_x(\Delta t)$.
Note that Eq.(\ref{C_OU}) represents the covariance at zero time lag, which is a special case of the covariance at time lag $\Delta t$, namely $Q=Q_x(0)$.
The covariance at finite time lag $\Delta t=t'-t''$ is equal to \cite{gardiner}

\begin{equation}\label{C_lag}
Q_x(\Delta t)=\overline{\Delta\textbf{x}(t')\Delta\textbf{x}(t'')^\dagger}=\left\{\begin{array}{ll}e^{(G-I)\Delta t}\;Q&\mbox{   for }\Delta t\geq 0\\Q\;
e^{-(G^\dagger-I)\Delta t}&\mbox{   for }\Delta t<0\end{array}\right.
\end{equation}
We define the temporally integrated activity as a linear convolution of the activity, namely

\begin{equation}
\textbf{y}(t)=\int_{-\infty}^{+\infty}dt'\;h(t-t')\textbf{x}(t')
\end{equation}where $h(t)$ is a given convolution kernel.
Using the Wiener-Khinchin theorem and the convolution theorem, it is straightforward to calculate the covariance of the integrated activity, which is equal to

\begin{equation}
Q_y(\Delta t)=\overline{\Delta\textbf{y}(t')\Delta\textbf{y}(t'')^\dagger}=\int_{-\infty}^{+\infty}dt\;h_2(\Delta t-t)Q_x(t)\end{equation}
where the second order kernel is equal to $h_2(t)=\int_{-\infty}^{+\infty} dt' h(t')h(t'+t)$.
If time integration is slow enough, such that the kernels $h$ and $h_2$ are approximately constant in a time interval in which the covariance $Q_x$ is sensibly different from zero, the above expression simplifies to

\begin{eqnarray*}
Q_y(\Delta t)\simeq h_2(\Delta t)\int_{-\infty}^{+\infty}dt\;Q_x(t)=h_2(\Delta t)\left[(I-G)^{-1}Q+Q(I-G^\dagger)^{-1}\right]=\\
=h_2(\Delta t)\overline{\Delta x_{ext}^2}\;(I-G)^{-1}G_{ext}G_{ext}^\dagger(I-G^\dagger)^{-1}
\end{eqnarray*}In the last two equalities we have, respectively, integrated Eq.(\ref{C_lag}) and used Eq.(\ref{lyapunov}), which we have multiplied by $(I-G)^{-1}$ on the
 left side and by $(I-G^\dagger)^{-1}$ on the right side.
The latter expression can be averaged over the network disorder to compute the mean correlation of the integrated activity.
We will consider only the case of $\Delta t=0$, since the case $\Delta t\neq 0$ is straightforward and is not our focus, and we denote $Q_y=Q_y(0)$. 
In addition, we substitute $h_2(0)=T^{-1}$, where $T$ is defined as the characteristic integration time of the kernel.

The on-diagonal elements of $Q_y$ are the temporal variances of the integrated activity of different neurons.
As in the computation of the variance of $x$, we take the sample mean across neurons and we average over the spatial disorder, obtaining
\begin{equation}
\left<\overline{\Delta y^2}\right>=\left<\frac{1}{N}\mbox{Tr}(Q_y)\right>=\frac{\overline{\Delta x_{ext}^2}}{TN}\left<\mbox{Tr}\left((I-G^\dagger)^{-1}(I-G)^{-1}G_{ext}G_{ext}^\dagger\right)\right>
\end{equation}
Again, we applied the trace operator to select the diagonal elements, we used its cyclic invariance, and since $G$ and $G_{ext}$ are independent we can average separately the factors involving the two matrices.
We use $\left<G_{ext}G_{ext}^\dagger\right>=k_{ext}g_{ext}^2\mathbf{1}\mathbf{1}^\dagger+\lambda_{ext}^2I$, and Eqs(\ref{appeq2},\ref{appeq3}) in Appendix 2, to obtain

\begin{equation}
\label{ondiagY}
\left<\overline{\Delta y^2}\right>=\frac{\overline{\Delta x_{ext}^2}}{T(1-\lambda^2)}\left[\frac{k_{ext}g_{ext}^2}{(1+g\sqrt{K})^2}+\lambda_{ext}^2\right]
\end{equation}
Note that the first term in square brackets is small, of order $K^{-1}$, and could be neglected. 

Next, we calculate the average covariance of the integrated activity, by looking at the off-diagonal elements of the matrix $Q_y$.
The off-diagonal elements are the pairwise covariances, and the sample mean across neuron pairs can be averaged over the spatial disorder to obtain the average covariance.
As in the computation of the covariance of $x$, we use the matrix $(\mathbf{1}\mathbf{1}^\dagger-I)$ to select the off-diagonal elements, and we obtain

\begin{eqnarray*}\left<\overline{\Delta y'\Delta y''}\right>&=&\left<\frac{1}{N(N-1)}\mbox{Tr}\left((\mathbf{1}\mathbf{1}^\dagger-I)Q_y\right)\right>=
\\&=&\frac{\overline{\Delta x_{ext}^2}}{TN(N-1)}\left<\mbox{Tr}\left((I-G^\dagger)^{-1}(\mathbf{1}\mathbf{1}^\dagger-I)(I-G)^{-1}G_{ext}G_{ext}^\dagger\right)\right>
\end{eqnarray*}
Again, we used the cyclic invariance of the trace operator and, since $G$ and $G_{ext}$ are independent, we can average separately the factors involving the two matrices.
We use $\left<G_{ext}G_{ext}^\dagger\right>=k_{ext}g_{ext}^2\mathbf{1}\mathbf{1}^\dagger+\lambda_{ext}^2I$ and Eqs.(\ref{appeq2},\ref{appeq3},\ref{appeq3b}) in Appendix 2.
Because the leading term is $K^{-1}$, here we keep terms of order $N^{-1}$, and we obtain

\begin{equation}
\label{offdiagY}
\left<\overline{\Delta y'\Delta y''}\right>=\frac{\overline{\Delta x_{ext}^2}}{T}\left[\frac{k_{ext}g_{ext}^2}{(1+g\sqrt{K})^2}-\frac{1}{N}\frac{\lambda_{ext}}{1-\lambda^2}\right]
\end{equation}

The mean correlation of the integrated activity is obtained by dividing the covariance, Eq.(\ref{offdiagY}), by the variance, Eq.(\ref{ondiagY}), i.e. (we assume that variance and covariance are independent):

\begin{equation}
\label{corrY}
\left<R_y\right>=\frac{\left<\overline{\Delta y'\Delta y''}\right>}{\left<\overline{\Delta y^2}\right>}=\frac{k_{ext}g_{ext}^2}{(1+g\sqrt{K})^2}\frac{(1-\lambda^2)}{\lambda_{ext}^2}-\frac{1}{N}
\end{equation}
Note that we neglected the term of order $K^{-1}$ in using Eq.(\ref{ondiagY}).
This expression shows that correlations of integrated activity can be negative and are small, of order $K^{-1}$.

\section*{Appendix 2: Traces of random matrix products}
In this section we introduce the diagrammatic notation to calculate the quenched averages of random matrix products (see e.g. \cite{novak03}).
In the context of neural networks, a diagrammatic notation has been also implemented recently by \cite{rangan09, pernice11, trousdale12}.
Theoretical results are obtained for the Gaussian distribution, although numerical simulations suggest that they generalize to other distributions with the same mean and variance (e.g. Bernouilli).
We will consider the case in which the mean of the matrix element is $\sim g/N$ and then recover the scaling studied in the main text by analytical continuation and the substitution $g\rightarrow g\sqrt{K}$.
We conclude the section by studying the case of non-homogeneous mean (e.g. interconnected excitatory and inhibitory neurons). 

\begin{figure}[h!]
\begin{center}
\includegraphics[width=4in]{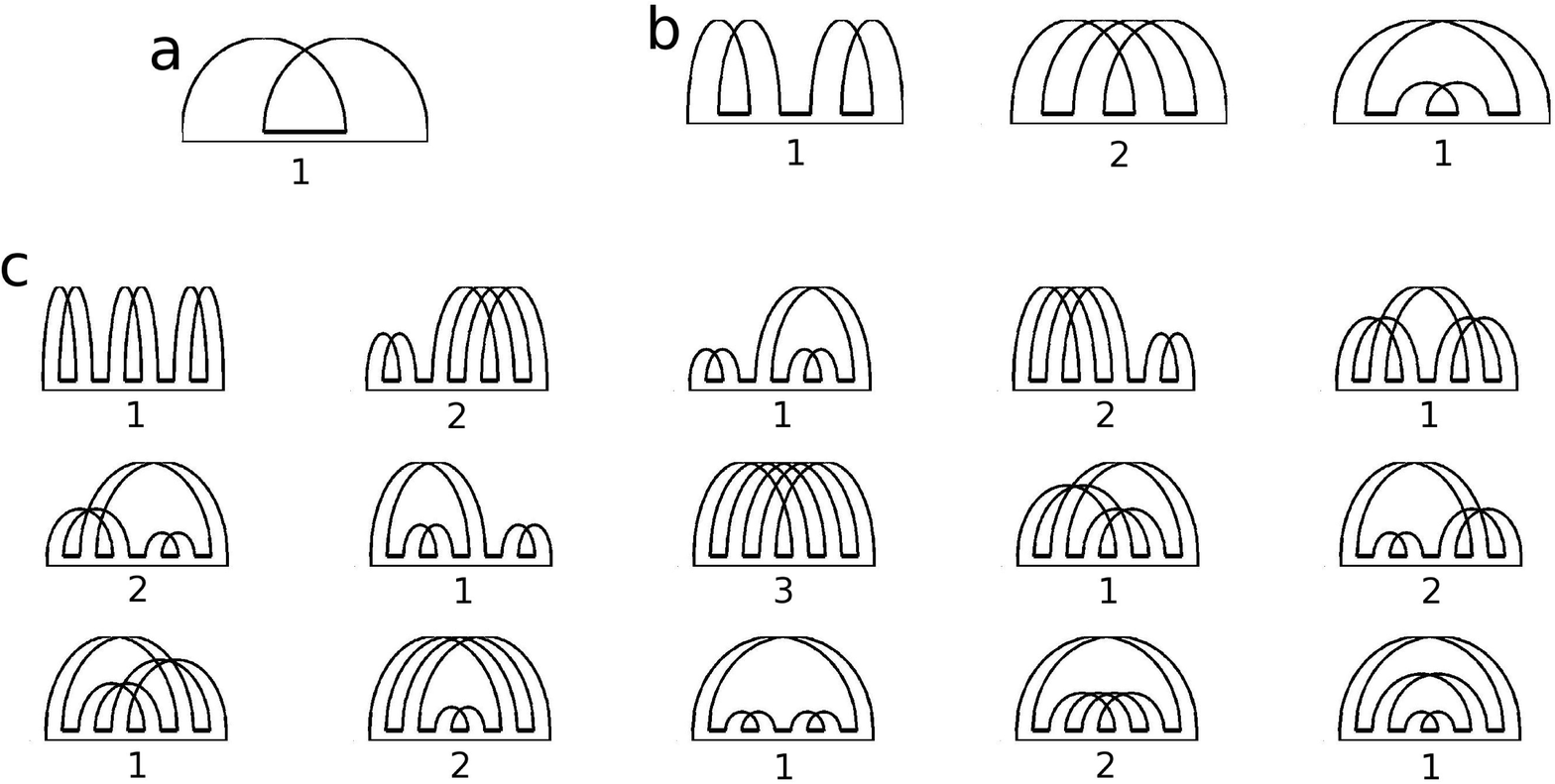}
\end{center}
\caption{
Diagrams of the traces of random matrix powers, described by Eq.(\ref{notranspose}). 
Below each diagram the number of its closed loops is indicated.
a) Second order. 
b) Fourth order.
c) Sixth order.}
\label{diag1}
\end{figure}

We start with the problem of calculating the quenched average of the trace of a power of the random matrix $R$ in the limit of large $N$ (where the size of the matrix is $N\times N$).
The matrix $R$ is characterized by independent and normally distributed elements, each element having zero mean and variance $N^{-1}$, namely

\begin{equation}
\left<R_{ij}\right>=0\;\;\;\;\;\;\;\left<R_{ij}^2\right>=\frac{1}{N}
\end{equation}
We start by calculating the second order, i.e. the average trace of the square of $R$.
For convenience of notation, we omit the sum over the indices (in this case the sum over the indices $a,b,c,d$)

\begin{equation}
\label{trR2}
\left<\mbox{Tr}\left(R^2\right)\right>=\delta_{ad}\delta_{bc}\left<R_{ab}R_{cd}\right>=N^{-1}\;\delta_{ad}\delta_{bc}\delta_{ac}\delta_{bd}
\end{equation}
The diagram corresponding to this expression is shown in Fig.\ref{diag1}a.
The diagram is obtained by drawing one node for each one of the four indices $a, b, c, d$, and by drawing an edge for each delta function in the expression, where the two nodes connected by the edge correspond to the two indices of the delta function.
Horizontal edges are due to the operations of trace (base edge) and matrix multiplication (middle edge), while arc-shaped edges are due to averaging.
The multiple edges determine different paths, and each pair of nodes connected by a path (even if not linked by an edge), corresponds to a pair of indices that must be equal, since they are connected by a sequence of delta functions.
Therefore, for each closed loop in the diagram there is one redundant delta function, which can be eliminated without performing the sum over the corresponding indices.
This implies that each closed loop contributes with a factor $N$, due to a free sum over $N$ elements.
Since the diagram for the second order has one loop, we have $\left<\mbox{Tr}\left(R^2\right)\right>=N^{-1}N=1$.

Note that all terms of odd order are zero, because $\left<R_{ij}^k\right>=0$ for odd $k$.
The next order is therefore the fourth order, which is equal to (again we omit the sum over all indices)
\begin{equation}
\left<\mbox{Tr}\left(R^4\right)\right>=\delta_{ah}\delta_{bc}\delta_{de}\delta_{fg}\left<R_{ab}R_{cd}R_{ef}R_{gh}\right>=
\end{equation}
\begin{equation}
=N^{-2}\;\delta_{ah}\delta_{bc}\delta_{de}\delta_{fg}\left[\delta_{ac}\delta_{bd}\delta_{eg}\delta_{fh}+\delta_{ae}\delta_{bf}\delta_{cg}\delta_{dh}+\delta_{ag}\delta_{bh}\delta_{ce}\delta_{df}\right]
\end{equation}
The fourth order has three diagrams, one for each term in the sum, shown in Fig.\ref{diag1}b.
The middle diagram has two closed loops, while the other two have only one loop.
Therefore the other two terms can be neglected, the middle term contributes with a factor $N^2$ and the fourth order gives $\left<\mbox{Tr}\left(R^4\right)\right>=1$.
The contribution of the fourth order moment ($\left<R^4_{ij}\right>$) can be neglected, because the corresponding terms in the sum have quartets of indices with the same value. 
The number of those terms is smaller by a factor of $N^2$ with respect to the number of second order terms. 
Similar arguments apply for higher order moments.

The fifteen diagrams of the sixth order are shown in Fig.\ref{diag1}c.
Again, we neglect moments higher than the second, and we note that only one diagram contributes with three loops, therefore $\left<\mbox{Tr}\left(R^6\right)\right>=1$.
By iterating this procedure, we find that order $2k$ has $(2k-1)!!$ diagrams of which only one has $k$ loops, therefore
\begin{equation}
\label{notranspose}
\left<\mbox{Tr}\left(R^{2k}\right)\right>=1
\end{equation}
for all values of $k$.

Note that the elements of the matrix $R$ have zero mean, while the matrices considered in the main text ($G$ and $G_{ext}$) have non-zero mean.  
As explained below, in order to calculate the average trace of matrix powers with non-zero mean, we need to compute averages where $R$ is interleaved by the matrix of ones.
We denote by $\mathbf{1}$ the column vector of $N$ components all equal to one, by $\mathbf{1}^\dagger$ the row vector, and by $\mathbf{1}\mathbf{1}^\dagger$ the $N\times N$ matrix with all elements equal to one (we denote by $\dagger$ the transpose operation).
We consider the two second order terms

\begin{equation}
\left<\mbox{Tr}\left(R\mathbf{1}\mathbf{1}^\dagger R\right)\right>=\delta_{ad}\left<R_{ab}R_{cd}\right>=N^{-1}\delta_{ad}\delta_{ac}\delta_{bd}=1
\end{equation}
\begin{equation}
\left<\mbox{Tr}\left(R^2\mathbf{1}\mathbf{1}^\dagger\right)\right>=\delta_{bc}\left<R_{ab}R_{cd}\right>=N^{-1}\delta_{bc}\delta_{ac}\delta_{bd}=1
\end{equation}
It is not surprising that these two expressions are equal, since the trace is cyclic invariant.
The only difference of these expressions with Eq.(\ref{trR2}) is the absence of a factor $\delta_{bc}$ in the former expression, and $\delta_{ad}$ in the latter.
This corresponds to cutting, respectively, the middle and the base horizontal edges in the diagram of Fig.\ref{diag1}a.
In general, inserting a matrix of ones at a given point of the sequence of $R$ products is equivalent to cutting the horizontal edge at that point in the corresponding diagram.
If the edge belongs to a closed loop, the cut has the only effect of removing a redundant delta function, and there is no change in the contribution of that diagram to the sum;
Conversely, if the edge belongs to an open path, the cut determines an additional $N$ factor, because the delta function removed was not redundant.
Since all diagrams have at least one closed loop, inserting a single matrix of ones has no effect at all orders.
Therefore,

\begin{equation}
\label{trR11R}
\left<\mbox{Tr}\left(R^{2k-k'}\mathbf{1}\mathbf{1}^\dagger R^{k'}\right)\right>=1
\end{equation}
for all $k'=0,\ldots,2k$.
Unless more loops are available to cut, inserting more matrices of ones may cut open paths, therefore the trace may be multiplied by $N$.
An additional $N$ factor is obtained also by multiplying the matrix of ones with itself, which occurs whenever additional matrices are inserted at same point in the sequence (we have that $\mathbf{1}^\dagger\mathbf{1}=N$ and $(\mathbf{1}\mathbf{1}^\dagger)^k=N^{k-1}\mathbf{1}\mathbf{1}^\dagger$ if $k>0$).

Using the above results, we can calculate the average trace of random matrix powers with non-zero mean and arbitrary variance (provided that the variance is of order $N^{-1}$).
We consider the matrix $G$ equal to

\begin{equation}
G=\frac{g}{N}\mathbf{1}\mathbf{1}^\dagger+\lambda R
\end{equation}
Note that the mean of this matrix has a different scaling with respect to that considered in the main text, but we will recover the latter by the substitution $g\rightarrow\;-\sqrt{K}g$.
A power of $G$ is calculated by multiplying $G$ to itself, and this determines an ordered product of powers of the matrices $R$ and $\mathbf{1}\mathbf{1}^\dagger$.
Note that these two matrices do not commute, therefore the binomial theorem cannot be applied.
We consider the average trace
\begin{equation}
\left<\mbox{Tr}\left(G^k\right)\right>=\sum_{k'=0}^kN^{-k'}g^{k'}\lambda^{k-k'}\sum^{{k \choose k'}}\left<\mbox{Tr}(\ldots)\right>
\end{equation}
where the trace in the right hand side is applied to an ordered product of $k'$ matrices $\mathbf{1}\mathbf{1}^\dagger$ and $k-k'$ matrices $R$, and the sum runs over all the ${k \choose k'}$ ordered products for a given $k$ and $k'$.
Using the above results, we find that the contribution of any of those traces is zero for $k-k'$ odd, is equal to one for $k'=0$ (provided that $k$ is even), is equal to $N^k$ for $k'=k$ and is at most of order $N^{k'-1}$ for $k'=1,\ldots,k-1$. 
Therefore, the leading order terms are $k'=k$ (for any value of $k$) and $k'=0$ (for $k$ even), all other terms can be neglected, and we find

\begin{equation}
\left<\mbox{Tr}\left(G^k\right)\right>=g^k+\lambda^k\delta_{k,even}
\end{equation}
If the matrix $G^k$ is further multiplied by a matrix of ones, the term $k'=0$ can also be neglected, and we find that

\begin{equation}
\left<\mbox{Tr}\left(G^k\mathbf{1}\mathbf{1}^\dagger\right)\right>=Ng^k=\mbox{Tr}\left(\left<G\right>^k\mathbf{1}\mathbf{1}^\dagger\right)
\end{equation}
for all values of $k$.
Note that if the mean of $G$ has a higher order in $N$, the result still holds.
This expression is particularly useful to compute the average of bracket expressions.
Because $\mbox{Tr}\left(A\mathbf{x}\mathbf{y}^\dagger\right)$=$\mathbf{y}^\dagger A\mathbf{x}$ for any matrix $A$ and vectors $\mathbf{x}$, $\mathbf{y}$, the expression can be rewritten as

\begin{equation}
\left<\mathbf{1}^\dagger G^k\mathbf{1}\right>=\mathbf{1}^\dagger \left<G\right>^k\mathbf{1}
\end{equation}
for all values of $k$.
Since any infinitely differentiable function $f$ can be expanded in Taylor series, the above result implies that

\begin{equation}
\left<\mathbf{1}^\dagger f\left(G\right)\mathbf{1}\right>=\mathbf{1}^\dagger f\left(\left<G\right>\right)\mathbf{1}
\end{equation}
Therefore, the following expression can be calculated and used to compute the mean activity in the main text,

\begin{equation}
\label{appeq1}
\left<\mathbf{1}^\dagger \left(I-G\right)^{-1}\mathbf{1}\right>=\frac{N}{1-g}
\end{equation}
Note that the substitution $g\rightarrow -g\sqrt{K}$ must be applied to recover the scaling studied in the main text.

\begin{figure}[h!]
\begin{center}
\includegraphics[width=4in]{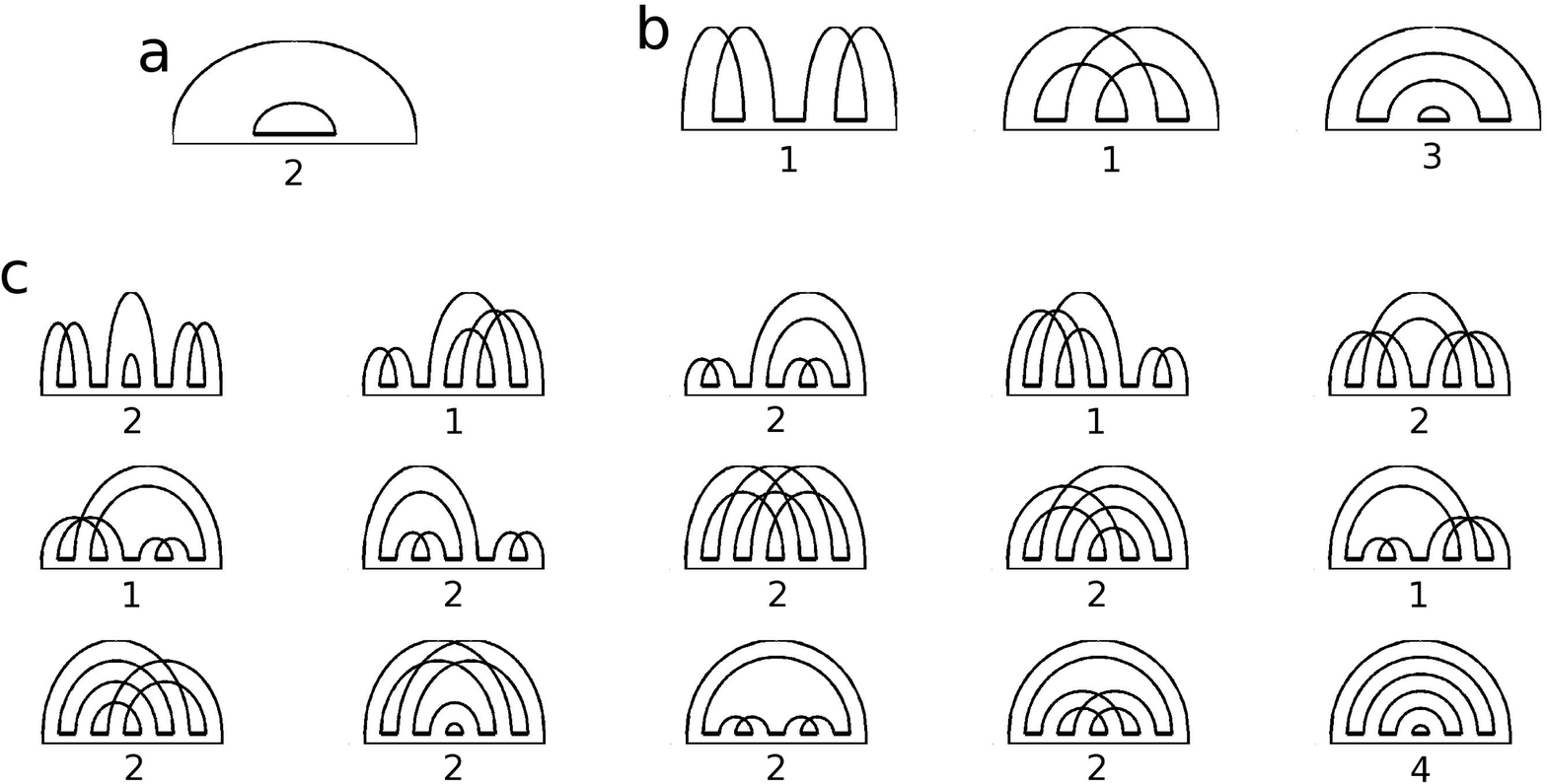}
\end{center}
\caption{
Diagrams of the traces of random matrix powers multiplied by its transpose, described by Eq.(\ref{yestranspose}). 
Below each diagram the number of its closed loops is indicated.
a) Second order. 
b) Fourth order.
c) Sixth order.}
\label{diag2}
\end{figure}

Next, we calculate the diagrammatic expansion for products of a random matrix with its transpose.
Again, all odd orders vanish and we neglect moments higher than the second at all orders.
The second order term is

\begin{equation}
\left<\mbox{Tr}\left(RR^\dagger\right)\right>=\delta_{ad}\delta_{bc}\left<R_{ab}R_{dc}\right>=N^{-1}\;\delta_{ad}\delta_{bc}\delta_{ad}\delta_{bc}
\end{equation}
The corresponding diagram has two loops and is shown in Fig.\ref{diag2}a, therefore the loops contribute with a factor $N^2$ and the second order is $\left<\mbox{Tr}\left(RR^\dagger\right)\right>=N$.
The fourth order is equal to

\begin{equation}
\left<\mbox{Tr}\left(R^2{R^2}^\dagger\right)\right>=\delta_{ah}\delta_{bc}\delta_{de}\delta_{fg}\left<R_{ab}R_{cd}R_{fe}R_{hg}\right>=
\end{equation}
\begin{equation}
=N^{-2}\;\delta_{ah}\delta_{bc}\delta_{de}\delta_{fg}\left[\delta_{ac}\delta_{bd}\delta_{fh}\delta_{eg}+\delta_{af}\delta_{be}\delta_{ch}\delta_{dg}+\delta_{ah}\delta_{bg}\delta_{cf}\delta_{de}\right]
\end{equation}
The three diagrams are shown in Fig.\ref{diag2}b.
The first two diagrams have one loop, while the third has three.
Therefore, that diagram contributes with a factor $N^3$ and the fourth order is equal to $\left<\mbox{Tr}\left(R^2{R^2}^\dagger\right)\right>=N$.
The diagrams for the sixth order are shown in Fig.\ref{diag2}c: only one diagram has four loops, and no diagram has three, therefore $\left<\mbox{Tr}\left(R^3{R^3}^\dagger\right)\right>=N$.
Iterating the procedure, we find that order $2k$ has $(2k-1)!!$ diagrams of which only one has $k+1$ loops, therefore
\begin{equation}
\label{yestranspose}
\left<\mbox{Tr}\left(R^k{R^k}^\dagger\right)\right>=N
\end{equation}
for all values of $k$.
Other combinations of powers of $R$ and its transpose give a smaller contribution, i.e. $\left<\mbox{Tr}\left(R^{2k-k'}{R^{k'}}^\dagger\right)\right>=o(1)$ for $k'\neq k$.

Inserting matrices of ones in this case has a similar effect as in the case above, Eq.(\ref{trR11R}), each matrix cuts the horizontal edge corresponding to where the matrix is placed.
Again, since each diagram has at least one loop, the insertion of a single matrix of ones (and the consequent edge removal) has no effect on the trace at all orders.
Therefore,

\begin{equation}
\left<\mbox{Tr}\left(R^{k-k'}\mathbf{1}\mathbf{1}^\dagger R^{k'}{R^{k}}^\dagger\right)\right>=\left<\mbox{Tr}\left(R^{k}{R^{k'}}^\dagger\mathbf{1}\mathbf{1}^\dagger{R^{k-k'}}^\dagger\right)\right>=N
\end{equation}
An insertion in a term with unequal powers of $R$ and $R^\dagger$ remains of order one.
Adding more matrices increases the trace by an order $N$ for each matrix, provided that no further loops are cutted.

Using the above expressions, we can compute the average of products of powers of the matrix $G$ and its transpose, namely

\begin{equation}
\left<\mbox{Tr}\left(G^k{G^{l}}^\dagger\right)\right>=\sum_{k'=0}^k\sum_{l'=0}^lN^{-k'-l'}g^{k'+l'}\lambda^{k+l-k'-l'}\sum^{{k \choose k'}{l \choose l'}}\left<\mbox{Tr}(\ldots)\right>
\end{equation}
where the trace in the right hand side is applied to an ordered product of $k'+l'$ matrices $\mathbf{1}\mathbf{1}^\dagger$, $k-k'$ matrices $R$ and $l-l'$ matrices $R^\dagger$.
If $k'=k$ and $l'=l$, the trace is equal to $N^{k+l}$, and the term is of order one.
If $k'=0$ and $l'=0$, the trace contributes with an order $N$, provided that $k=l$.
If $k-k'=l-l'$, the traces contribute at most with an order $N^{k'+l'}$, and the term is of order one, while if $k-k'\neq l-l'$ the term is of smaller order.
Therefore the leading order is $N$, and we have

\begin{equation}
\left<\mbox{Tr}\left(G^k{G^{l}}^\dagger\right)\right>=N\delta_{kl}\lambda^{k+l}
\end{equation}
In the case in which matrices of ones are inserted, the term $k'=0$, $l'=0$ is no longer leading, and many other terms have to be considered.
Those are the terms for $k-k'=l-l'$, and for which additional inserted matrices cuts the same loop.
Since the leading diagrams at all orders have one two-nodes loop in the middle and one at the boundaries, if a matrix of ones is inserted in the middle or at the boundaries, additional matrices must continue to be inserted at the same place in order to cut the same loop.
We eliminate one sum and we use the index $m=k-k'=l-l'$ in place of $k'$ and $l'$;
We obtain

\begin{equation}
\left<\mbox{Tr}\left(G^k{G^{l}}^\dagger\mathbf{1}\mathbf{1}^\dagger\right)\right>=\sum_{m=0}^{\min{(k,l)}}N^{2m-k-l}g^{l+k-2m}\lambda^{2m}\left<\mbox{Tr}\left((\mathbf{1}\mathbf{1}^\dagger)^{k-m}R^m{R^m}^\dagger(\mathbf{1}\mathbf{1}^\dagger)^{l-m+1}\right)\right>
\end{equation}

\begin{equation}
\left<\mbox{Tr}\left(G^k\mathbf{1}\mathbf{1}^\dagger{G^{l}}^\dagger\right)\right>=\sum_{m=0}^{\min{(k,l)}}N^{2m-k-l}g^{l+k-2m}\lambda^{2m}\left<\mbox{Tr}\left(R^m(\mathbf{1}\mathbf{1}^\dagger)^{k+l-2m+1}{R^m}^\dagger\right)\right>
\end{equation}
Both expressions are equal to

\begin{equation}
\left<\mbox{Tr}\left(G^k{G^{l}}^\dagger\mathbf{1}\mathbf{1}^\dagger\right)\right>=\left<\mbox{Tr}\left(G^k\mathbf{1}\mathbf{1}^\dagger{G^{l}}^\dagger\right)\right>=N\sum_{m=0}^{\min{(k,l)}}g^{l+k-2m}\lambda^{2m}
\end{equation}
Furthermore, we calculate the average trace with two inserted matrices.
In that case, the leading term is for $k=k'$ and $l=l'$ (or $m=0$), and we obtain

\begin{equation}
\left<\mbox{Tr}\left(G^k\mathbf{1}\mathbf{1}^\dagger{G^{l}}^\dagger\mathbf{1}\mathbf{1}^\dagger\right)\right>=N^2\;g^{k+l}=\mbox{Tr}\left(\left<G\right>^k\mathbf{1}\mathbf{1}^\dagger{\left<G\right>^{l}}^\dagger\mathbf{1}\mathbf{1}^\dagger\right)
\end{equation}

Using the expressions above and the Taylor series expansion of infinitely differentiable functions, we calculate the following traces that are used in Appendix 1 to compute the variance and covariance of the activity

\begin{equation}
\label{appeq2}
\left<\mbox{Tr}\left((I-G)^{-1}(I-G^\dagger)^{-1}\right)\right>=\frac{N}{1-\lambda^2}
\end{equation}

\begin{equation}
\label{appeq3}
\left<\mbox{Tr}\left((I-G)^{-1}\mathbf{1}\mathbf{1}^\dagger(I-G^\dagger)^{-1}\right)\right>=\frac{N}{(1-\lambda^2)(1-g)^2}
\end{equation}

\begin{equation}
\label{appeq3b}
\left<\mbox{Tr}\left((I-G)^{-1}\mathbf{1}\mathbf{1}^\dagger(I-G^\dagger)^{-1}\mathbf{1}\mathbf{1}^\dagger\right)\right>=\frac{N^2}{(1-g)^2}
\end{equation}

\begin{equation}
\label{appeq4}
\int_0^\infty dt\;e^{-2t}\left<\mbox{Tr}\left(e^{Gt}e^{G^\dagger t}\right)\right>=\frac{N}{2\sqrt{1-\lambda^2}}
\end{equation}

\begin{equation}
\label{appeq5}
\int_0^\infty dt\;e^{-2t}\left<\mbox{Tr}\left(e^{Gt}\mathbf{1}\mathbf{1}^\dagger e^{G^\dagger t}\right)\right>=\frac{N}{2\sqrt{1-\lambda^2}\left(1-g\right)}\left[\frac{1+\sqrt{1-\lambda^2}\left(1-g\right)}{1+\sqrt{1-\lambda^2}-g}\right]
\end{equation}

\begin{equation}
\label{appeq6}
\int_0^\infty dt\;e^{-2t}\left<\mbox{Tr}\left(e^{Gt}\mathbf{1}\mathbf{1}^\dagger e^{G^\dagger t}\mathbf{1}\mathbf{1}^\dagger\right)\right>=\frac{N^2}{2(1-g)}
\end{equation}
Note that the substitution $g\rightarrow -g\sqrt{K}$ must be applied to recover the scaling studied in the main text. 
If $K$ is proportional to $N$, this substitution may change the order of magnitude of various terms in the summation considered above, possibly modifying the leading terms in each sum. 
Note that all series converge only for $|g|<1$, but their sum can be evaluated at $g\rightarrow -g\sqrt{K}$ by analytical continuation.
Then, approximating the sums by the leading terms described above is accurate under the assumption that all series involving lower order terms converge to bounded functions of $g$. 

We conclude this section by studying the case of non-homogeneous mean and variance.
Above, we have assumed that the mean and variance are homogeneous, namely they take the same value for different matrix elements: $\left<G_{ij}\right>=g/N$ and $\left<\Delta G_{ij}^2\right>=\lambda^2/N$.
However, the same methods could be used to analyze the more general case in which the mean and variance are inhomogeneous.
In fact, as long as the mean and variances do not depend on $N$, they do not change the order of different terms in the sums considered above.
Therefore, the calculation would consist in taking only the leading terms and recalculate their value according to the new matrices of means and variances.
For example, even in the inhomogeneous case, the sums resulting in Eqs.(\ref{appeq1}),(\ref{appeq3b}),(\ref{appeq6}) would be still determined uniquely by the mean $\left<G_{ij}\right>$, and the sums resulting in Eqs.(\ref{appeq2}),(\ref{appeq4}) would be still determined uniquely by the variance $\left<\Delta G_{ij}^2\right>$.
Sums affected by both the mean and variance, such as those resulting in Eqs.(\ref{appeq3}),(\ref{appeq5}), would still be calculated by using only the leading terms determined above. 

A particularly simple case is when the mean and variance depend only on the pre-synaptic neuron, i.e. $\left<G_{ij}\right>=g_j/N$ and $\left<\Delta G_{ij}^2\right>=\lambda_j^2/N$.
This includes the case of interconnected excitatory and inhibitory neurons, where $g_j$ is positive for excitatory neurons and negative for inhibitory neurons.
In that case, all results above still hold, with the simple substitutions

\begin{equation}
g\longleftarrow N^{-1}\sum_{j=1}^N g_j
\end{equation}

\begin{equation}
\lambda^2\longleftarrow N^{-1}\sum_{j=1}^N \lambda_j^2
\end{equation}
Namely, the parameters $g$ and $\lambda^2$ now measures the mean connection strength and the mean variance across pre-synaptic neurons.
Simulations suggest that a similar substitution, a mean of $g$ and $\lambda^2$ across all matrix entries, works well even for general non-homogeneous parameters.

\end{document}